\documentclass[aps,prb,twocolumn,showpacs,preprintnumbers,superscriptaddress]{revtex4}
\usepackage{bm,amsmath,amssymb,color}
\usepackage[mathcal]{euscript}
\usepackage[hypertex]{hyperref}
\usepackage{graphicx}
\usepackage{subfig}
\renewcommand{\vec}[1]{\bm{#1}}
\begin{document}

\title{Effective anisotropy of thin nanomagnets: beyond the surface anisotropy approach}

\author{Jean--Guy Caputo}
 \affiliation{Laboratoire de Math\'ematiques, INSA de Rouen, B.P. 8, 76131 Mont-Saint-Aignan cedex, France}
 \affiliation{Laboratoire de Physique theorique et modelisation, Universit\'e
 de Cergy-Pontoise and C.N.R.S., Cergy-Pontoise, France}

\author{Yuri Gaididei}
 \affiliation{Institute for Theoretical Physics, 03143 Kiev, Ukraine}

\author{Volodymyr P. Kravchuk}
 \affiliation{National Taras Shevchenko University of Kiev, 03127 Kiev, Ukraine}

\author{Franz G.~Mertens}
 \affiliation{Physics Institute, University of Bayreuth, 95440 Bayreuth, Germany}

\author{Denis D. Sheka}
 \email[Corresponding author. Electronic address:\\]{denis\_sheka@univ.kiev.ua}
 \affiliation{National Taras Shevchenko University of Kiev, 03127 Kiev, Ukraine}

\date{28.05.07}

%
%

\begin{abstract}
We study the effective anisotropy induced in thin nanomagnets by the nonlocal
demagnetization field (dipole-dipole interaction). Assuming a magnetization
independent of the thickness coordinate, we reduce the energy to an
inhomogeneneous onsite anisotropy. Vortex solutions exist and are ground
states for this model. We illustrate our approach for a disk and a square
geometry. In particular, we obtain good agreement between spin--lattice
simulations with this effective anisotropy and micromagnetic simulations.

\end{abstract}




\maketitle

\section{Introduction}
\label{sec:introduction}

Magnetic nanoparticles and structures have recently attracted a growing
interest for their physical properties and a number of possible applications.
\cite{Hubert98,Skomski03,Bader06} For example the vortex (ground) state of a
disk--shaped nanoparticle could provide high density storage and high speed
magnetic RAM \cite{Cowburn02}. The theoretical models for these systems have
been known for some time \cite{Brown63,Aharoni96} and include the nonlocal
demagnetization field. At microscopic level this field is due to the dipolar
interaction
\begin{equation} \label{eq:H-dipolar}
\!\!\!\!\! \mathcal{H}_{\text{d}} \!=\! \frac{D}{2}\!\!
\sum_{\substack{\vec{n}, \vec{m}\\\vec{n}\neq \vec{m}}}\! \left[
\frac{\vec{S}_{\vec{n}}\cdot \vec{S}_{\vec{m}}}{r_{\vec{n} \vec{m}}^3} - 3
\frac{ \left(\vec{S}_{\vec{n}}\cdot \vec{r}_{\vec{n} \vec{m}} \right)
\left(\vec{S}_{\vec{m}}\cdot \vec{r}_{\vec{n} \vec{m}} \right)}{r_{\vec{n}
\vec{m}}^5} \right]\!.
\end{equation}
Here $\vec{S}_{\vec{n}}\equiv\left(S^x_{\vec{n}}, S^y_{\vec{n}},
S^z_{\vec{n}}\right)$ is a classical spin vector with fixed length $S$ on the
site $\vec{n}=(n_x,n_y, n_z)$ of a three--dimensional lattice. The summation
runs over all magnets $(\vec{n},\vec{m})$, and $\vec{r}_{\vec{m}\vec{n}}
\equiv \vec{r}_{\vec{n}} - \vec{r}_{\vec{m}}$. The parameter $D = \mu_B^2 g^2$
is the strength of the long range dipolar interaction and $g$ is the
Lande--factor.

In the past analytical studies have been mainly limited to assuming a
homogeneous demagnetization field distribution, uniform Stoner--Wohlfarth
theory\cite{Stoner48} and near--uniform Brown's linear analysis
\cite{Skomski03}. Recent advances in nanotechnology and computing power
established the complexity of magnetization distribution in nanoparticles. For
example square nanoparticles exhibit buckling states, flower states, apple
states, leaf states etc, \cite{Usov92,Cowburn98,Ivanov04} when their size
exceeds the single--domain limit. In disk--shaped particles vortex states
\cite{Usov94,Hubert98}, edge fractional vortex states etc
\cite{Kireev03,Tchernyshyov05} appear. Some of these complex states can be
obtained by a small perturbation of a homogeneous state. For example
\citet{Cowburn98a} showed that dipolar interactions cause flower and leaf
states in square nanoparticles, which was confirmed by direct
experiments\cite{Cowburn98}. However the linear analysis does not work for
topologically nontrivial states like kinks, vortices etc. One possibility to
study these structures in nanomagnets is the Ritz variational method. It was
applied to analyze the vortex structure of the disk--shaped
nanodot\cite{Usov94,Hubert98}. A disadvantage of this method is to limit the
solution to a certain class of minimizers, so that one can usually study only
one type of excitation. Linear waves are left out together with their coupling
to the main excitation.

The various regimes were studied in
Refs.~\onlinecite{Gioia97,Desimone95,Desimone02,Kohn05,Kohn05a,Moser04,Kurzke06}.
The important length scale is the magnetic exchange length
$\ell=\sqrt{\textsf{A}/4\pi M_S^2}$ where $\textsf{A}$ is the exchange
constant and $M_S$ is the saturation magnetization. Depending on the relation
between the film diameter $2R$, its thickness $h$ and $\ell$ many scaling
limits can be analyzed, see Ref.~\onlinecite{Kohn05a} for an overview.
Probably the first rigorous study was made by \citet{Gioia97} who showed that
for an infinitesimally thin--film ($h/R\to0$, $\ell/R\to\text{const}$) the
magnetostatic energy tends to an effective 2D easy--plane anisotropy energy.
In this case the ground state is a homogeneous in--plane magnetization
state\cite{Gioia97}. This effective easy--plane anisotropy has a simple
magnetostatic interpretation. The sources of magnetostatic field are volume
and surface magnetostatic charges. For thin structures one can neglect the
volume charges. Face surface charges contribute to the energy density as $2\pi
M_z^2$ which is the same term one would get with an effective easy--plane
anisotropy\cite{Ivanov05}. In the case $h/R \ll 1$ and $\ell^2\ll 2
h\,R\,|\ln(h/2R)|$ the magnetization develops edge defects, including
fractional vortices.\cite{Moser04,Tchernyshyov05,Kurzke06} This problem has a
boundary constraint and an interior penalty. It is relevant for typical
Permalloy ($\text{Ni}_{80}\text{Fe}_{20}$, Py) disks where we have
$h\sim20$nm, $2R\sim100$nm and $\ell\sim5.3$nm.

It was shown in Refs.~\onlinecite{Kohn05,Kohn05a} that in the limit
$h/R\rightarrow 0$ under the scaling
\begin{equation} \label{eq:Kohn-limit}
\frac{2hR}{\ell^2}\left|\ln\frac{h}{2R}\right|\rightarrow C
\end{equation}
the full $3D$ micromagnetic problem reduces to a much simpler $2D$ variational
problem where the magnetostatic energy tends to the effective surface
anisotropy term
\begin{equation} \label{eq:surf}
E_{\text{surf}}=\int\limits_{\partial \Omega}(\vec{S}\cdot\vec{\tau})^2
\mathrm{d}S
\end{equation}
where $\vec{\tau}$ is the local tangent vector on the domain boundary
$\partial \Omega$. In this case the magnetization $\vec{S}$ has no out of
plane component $(S_z=0)$ and does not develop walls and vortices.

To study nanomagnets with curling ground states, here we develop a new
analytical approach. We split the dipole-dipole spin interaction
(\ref{eq:H-dipolar}) into two parts. The first one is an on-site anisotropy
with spatially dependent anisotropy coefficients. The second part represents
an effective dispersive interaction. The anisotropy interaction consists of
two terms: an easy-plane anisotropy and an in-plane anisotropy. We show that
the vortex state minimizes the effective in-plane anisotropy. We also show
that for ultra-thin nanomagnets $(h/R\to 0)$ the in-plane anisotropy term
reduces to the surface anisotropy \eqref{eq:surf}. For the nonhomogeneous
state which is our main interest, our approach is valid if
\begin{equation} \label{eq:effective-correct}
R\gg h \quad \text{and} \quad R\gg\ell.
\end{equation}

In Sec.~\ref{sec:eff-anis} we introduce our discrete model together with the
dipolar energy and adapt it to the plain--parallel spin--field distribution,
which is our main simplification. We further simplify the model by considering
only the local part of the dipolar energy, which results in an effective
anisotropy. In the continuum approximation of the system we get a local energy
functional with nonhomogeneous anisotropy coefficients, see
Sec.~\ref{sec:continuum}. The dispersive interaction is discussed in
Sec.~\ref{sec:disp}. To illustrate our method of effective anisotropy we
consider in Sec.~\ref{sec:disk} the disk--shape nanoparticle and study its
ground state spin distribution. Our simple model describes exactly the
homogeneous state (see Sec.~\ref{sec:disk-hom}) and very precisely the vortex
state (see Sec.~\ref{sec:disk-vortex}). In Sec.~\ref{sec:simulations} we
confirm our analysis by numerical simulations. These are done first for the
disk--shaped nanoparticle (Sec.~\ref{sec:disk-simulations}) and then for the
prism--shaped one (Sec.~\ref{sec:prism}). We discuss our results in
Sec.~\ref{sec:discussion}.

\section{Model. Effective anisotropy}
\label{sec:eff-anis} %

We consider a ferromagnetic system described by the classical Heisenberg
isotropic exchange Hamiltonian
\begin{equation} \label{eq:H-ex}
\mathcal{H}_{\text{ex}} =-\frac{J}{2}\!\! \sum_{\langle
\vec{n},\vec{n}'\rangle}\!\vec{S}_{\vec{n}} \vec{S}_{\vec{n}'} ,
\end{equation}
where the exchange integral $J > 0$ and the summation runs over
nearest--neighbor pairs $\langle \vec{n},\vec{n}'\rangle$. The total
Hamiltonian is a sum of the exchange energy \eqref{eq:H-ex} and the dipolar
one \eqref{eq:H-dipolar}.

Our main assumption is that $\vec{S}_{\vec{n}}$ depends only on the $x$ and
$y$ coordinates. Such a plane--parallel spin distribution is adequate for thin
films with a constant thickness $h=N_z a_0$, ($a_0$ being the lattice
constant) and nanoparticles with small aspect ratio. The exchange interaction
can be written as the sum of an intra--plane
$\mathcal{H}_{\text{ex}}^{\text{intra}}$ term and an inter--plane one
$\mathcal{H}_{\text{ex}}^{\text{inter}}$
\begin{equation} \label{eq:H-ex2}
\begin{split}
\mathcal{H}_{\text{ex}}^{\text{intra}} =&-\frac{(N_z+1)J}{2}\!\!
\sum_{\substack{\langle {\vec{\nu}},{\vec{\nu}}'\rangle}}\!
\vec{S}_{{\vec{\nu}}} \vec{S}_{{\vec{\nu}}'}, \\
\mathcal{H}_{\text{ex}}^{\text{inter}} =&-N_z J
\sum_{{\vec{\nu}}}\!\left(\vec{S}_{{\vec{\nu}}}\right)^2 = -N_z N_x N_y J S^2.
\end{split}
\end{equation}
Here and below the Greek index ${\vec{\nu}}=(n_x,n_y)$ corresponds to the XY
components of the vector $\vec{n}=(n_x,n_y,n_z)$. One can see that the
inter--plane interaction is equivalent to an on--site anisotropy. The
inter--exchange term gives a constant contribution, so it can be omitted.

Let us consider the dipolar energy. Using the above mentioned assumption about
the plane--parallel spin distribution, the dipolar Hamiltonian can be written
as (see Appendix \ref{sec:appendix-discrete} for the details):
\begin{subequations}
\begin{equation} \label{eq:H-dipolar-via-ABC}
\begin{split}
&\mathcal{H}_{\text{d}} = -\frac{D}{2}\!\!
\sum_{\substack{{\vec{\nu}},{\vec{\mu}}}}\! \Bigl[
A_{{\vec{\mu}}{\vec{\nu}}}\left(\vec{S}_{{\vec{\nu}}}\cdot
\vec{S}_{{\vec{\mu}}} - 3
S_{{\vec{\nu}}}^zS_{{\vec{\mu}}}^z\right)\\
&+ B_{{\vec{\mu}}{\vec{\nu}}} \left( S_{{\vec{\nu}}}^x S_{{\vec{\mu}}}^x -
S_{{\vec{\nu}}}^y S_{{\vec{\mu}}}^y \right) + C_{{\vec{\mu}}{\vec{\nu}}}\left(
S_{{\vec{\nu}}}^x S_{{\vec{\mu}}}^y + S_{{\vec{\nu}}}^y S_{{\vec{\mu}}}^x
\right)\Bigr].\end{split}
\end{equation}
\end{subequations}
Here the sum runs only over the 2D lattice XY. All the information about the
original 3D structure of our system is in the coefficients
$A_{{\vec{\mu}}{\vec{\nu}}}$, $B_{{\vec{\mu}}{\vec{\nu}}}$ and
$C_{{\vec{\mu}}{\vec{\nu}}}$,
\begin{subequations} \label{eq:A-B-C}
\begin{align}
\label{eq:A-B-C-(1)} %
A_{{\vec{\mu}}{\vec{\nu}}}&=\frac12
\sum_{\substack{m_z,n_z\\r_{\vec{n}\vec{m}}\neq0}}
\frac{r_{\vec{m}\vec{n}}^2 - 3z_{\vec{m}\vec{n}}^2}{r_{\vec{m}\vec{n}}^5}, \\ %
\label{eq:A-B-C-(2)} %
B_{{\vec{\mu}}{\vec{\nu}}}&= \frac32
\sum_{\substack{m_z,n_z\\r_{\vec{n}\vec{m}}\neq0}}
\frac{x_{\vec{m}\vec{n}}^2 - y_{\vec{m}\vec{n}}^2}{r_{\vec{m}\vec{n}}^5}, \\ %
\label{eq:A-B-C-(3)} %
C_{{\vec{\mu}}{\vec{\nu}}}&= 3
\sum_{\substack{m_z,n_z\\r_{\vec{n}\vec{m}}\neq0}} \frac{x_{\vec{m}\vec{n}}
y_{\vec{m}\vec{n}}}{r_{\vec{m}\vec{n}}^5}.
\end{align}
\end{subequations}

To gain insight into the anisotropic properties of the system we represent the
dipolar energy \eqref{eq:H-dipolar-via-ABC} as a sum
\begin{equation*}
\mathcal{H}_{\text{d}}=\mathcal{H}_{\text{d}}^{\text{loc}}+
\Delta\mathcal{H}_{\text{d}},
\end{equation*}
where
\begin{equation} \label{eq:Hd-loc}
\begin{split}
\mathcal{H}_{\text{d}}^{\text{loc}} &= -\frac{D}{2}\!\! \sum_{{\vec{\nu}}}\!
\Biggl\{ \bar{A}_{{\vec{\nu}}}\Bigl[\left(\vec{S}_{{\vec{\nu}}}\right)^2 -
3\left(
S_{{\vec{\nu}}}^z\right)^2\Bigr]\\
& + \bar{B}_{{\vec{\nu}}} \Bigl[\left(S_{{\vec{\nu}}}^x\right)^2 -
\left(S_{{\vec{\nu}}}^y\right)^2 \Bigr] + 2\bar{C}_{{\vec{\nu}}}
S_{{\vec{\nu}}}^x S_{{\vec{\nu}}}^y\Biggr\}.
\end{split}
\end{equation}
is an effective on--site anisotropic energy and
\begin{equation} \label{eq:Hd-disp}
\begin{split}
\Delta\mathcal{H}_{\text{d}} &= \frac{D}{4}\!\!
\sum_{\substack{{\vec{\nu}},{\vec{\mu}}}}\! \Biggl\{
A_{{\vec{\mu}}{\vec{\nu}}}\left[\left(\vec{S}_{{\vec{\nu}}}-
\vec{S}_{{\vec{\mu}}}\right)^2 - 3\,\left( S_{{\vec{\nu}}}^z
- S_{{\vec{\mu}}}^z\right)^2\right]\\
&+ B_{{\vec{\mu}}{\vec{\nu}}} \left[ \left(S_{{\vec{\nu}}}^x
-S_{{\vec{\mu}}}^x\right)^2 - \left(S_{{\vec{\nu}}}^y
-S_{{\vec{\mu}}}^y\right)^2 \right] \\
&+ 2C_{{\vec{\mu}}{\vec{\nu}}} \left(S_{{\vec{\nu}}}^x
- S_{{\vec{\mu}}}^x\right) \left(S_{{\vec{\nu}}}^y
-S_{{\vec{\mu}}}^y \right)\Biggr\}.
\end{split}
\end{equation}
is the dispersive part of the dipolar interaction. Here we introduce the
coefficients of \emph{effective anisotropy}
\begin{equation} \label{eq:A-B-C-discrete}
\begin{split}
\bar{A}_{{\vec{\nu}}} &= \sum_{{\vec{\mu}}}A_{{\vec{\mu}}{\vec{\nu}}},\;
\bar{B}_{{\vec{\nu}}} = \sum_{{\vec{\mu}}}B_{{\vec{\mu}}{\vec{\nu}}},\;
\bar{C}_{{\vec{\nu}}} = \sum_{{\vec{\mu}}}C_{{\vec{\mu}}{\vec{\nu}}}.
\end{split}
\end{equation}
The dipolar energy $\mathcal{H}_{\text{d}}^{\text{loc}}$ contains only local
interaction; it has a form of the anisotropy energy with nonhomogeneous
$\bar{A}_{{\vec{\nu}}}$, $\bar{B}_{{\vec{\nu}}}$, $\bar{C}_{{\vec{\nu}}}$. In
next sections we discuss these quantities. For this end we need to obtain the
continuum limit of our model.

\subsection{Continuum description}
\label{sec:continuum}

The continuum description of the system is based on smoothing the lattice
model, using the normalized magnetization
\begin{equation} \label{eq:S-via-M}
\vec{m}(\vec{r}) = \frac{g\mu_B}{a_0^3M_S} \sum_{\vec{n}} \vec{S}_{\vec{n}}
\delta(\vec{r} - \vec{r}_{\vec{n}}),
\end{equation}
where $M_S$ is the saturation magnetization. The exchange energy, the
continuum version of \eqref{eq:H-ex2} is
\begin{equation} \label{eq:E-exchange}
\begin{split}
\mathcal{E}_{\text{ex}} &= \tfrac12 \textsf{A} (h+a_0)\int \mathrm{d}^2x
\left(\vec\nabla \vec{m}\right)^2,
\end{split}
\end{equation}
where $\textsf{A}=JM_S^2 a_0^5/D$ is the exchange constant.

\begin{figure}
\includegraphics[width=0.7\columnwidth]{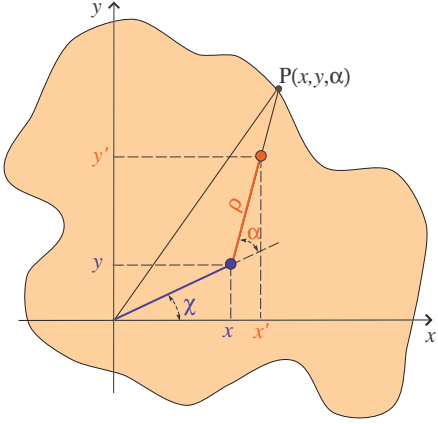}
\caption{(Color online) Arrangement of coordinates in the local reference
frame.}\label{fig:arb_shape}
\end{figure}

Now let us consider the dipolar energy and use its approximate Hamiltonian
\eqref{eq:Hd-loc}. To present this energy in a standard phenomenological form
one needs to transform the summation over the lattice to an integration over
the volume. There is a singularity for $\vec{r}_{\vec{m}\vec{n}} \to 0$. Using
a regularization similar to the one in Ref.~\onlinecite{Akhiezer68}, we find
(see Appendix \ref{sec:appendix-continuum} for details) that the local part of
the dipolar energy is
\begin{equation} \label{eq:E-dd-eff}
\begin{split}
\mathcal{E}_d &= \pi M_S^2 h \int \mathrm{d}^2 x \Biggl\{ \mathcal{A}(x,y)
\left[ 1 - 3
\cos^2\theta \right]\\
& + \sin^2\!\theta\ \text{Re}\Bigl[\mathcal{B}(x,y)
e^{2\imath(\phi-\chi)}\Bigr] \Biggr\},
\end{split}
\end{equation}
where we used the angular parameterization for the magnetization:
$m^z=\cos\theta$ and $m^x+im^y=\sin\theta e^{\imath\phi}$. Here and below we
dropped the \emph{loc} superscript. One can see that the original nonlocal
dipolar interaction results in an effective local anisotropy energy. The
coefficients of effective anisotropy $\mathcal{A}$ and $\mathcal{B}$ are
nonhomogeneous:
\begin{subequations} \label{eq:A-and-B}
\begin{align}
\label{eq:A-fin} %
\mathcal{A}(x,y) &= -\frac{2}{3} - \frac{a_0}{12h}\Biggl[8 \Theta_+(h) + 3
+ \frac{3a_0^3}{(a_0^2+h^2)^{3/2}} \Biggr] \nonumber\\
&+\frac{1}{2\pi}\int_0^{2\pi}\mathrm{d}\alpha \Biggl[
\frac{\sqrt{P^2+h^2}-P}{h} + \frac{a_0}{\sqrt{P^2+h^2}} \nonumber\\
& + \frac{a_0^2}{4Ph}+ \frac{a_0^2P^2}{4h(P^2+h^2)^{3/2}}\Biggr],\\
\label{eq:B-fin} %
\mathcal{B}(x,y) &= \frac{1}{2\pi}\int_0^{2\pi}
\mathcal{F}(P,h) e^{-2\imath\alpha}\mathrm{d}\alpha, \nonumber\\
\mathcal{F}(P,h) &= \frac{P-\sqrt{P^2+h^2}}{h}
- 2\left(1+\frac{a_0}{h}\right) \ln\frac{\sqrt{P^2+h^2}-h}{P} \nonumber \\
&+ \frac{a_0}{\sqrt{P^2+h^2}} + \frac{3a_0^2}{4Ph}
+ \frac{a_0^2}{4h}\frac{3P^2+2h^2}{(P^2+h^2)^{3/2}},
\end{align}
\end{subequations}
where the Heaviside function $\Theta_+(x)$ takes the unit values for any
positive $x$ and zero values for $x\leq0$.  In Eqs.~\eqref{eq:A-and-B} we used
the local reference frame
\begin{equation} \label{eq:loc-frame}
x' = x + \rho\cos(\chi+\alpha),\quad y' = y + \rho\sin(\chi+\alpha),
\end{equation}
which is centered at $(x,y)$. The term $P$ is the distance from this point to
the border of the system, it depends on the azimuthal angle $\alpha$ and
position $(x,y)$, see Fig.~\ref{fig:arb_shape}.

In the limit case of the pure 2D system (monolayer with $h=0$) the total
energy, normalized by the 2D area $\mathcal{S}$, takes a form
\begin{equation} \label{eq:W-mono}
\begin{split}
& W^{h=0} \equiv \frac{\mathcal{E}_{\text{ex}} +
\mathcal{E}_d}{M_S^2\mathcal{S}a_0} = W_{\text{ex}}^{h=0} + W_d^{h=0},\\
& W_{\text{ex}}^{h=0} = \frac{2\pi\ell^2}{\mathcal{S}}\int \mathrm{d}^2x
\left[ (\vec{\nabla}\theta)^2 +\sin^2\theta (\vec{\nabla}\phi)^2\right],\\
&W_d^{h=0} = \frac{\pi}{\mathcal{S}}\int \mathrm{d}^2x\Biggl\{
\mathcal{A}^{h=0}(x,y)\left[ 1 - 3 \cos^2\theta \right]\\
& + \sin^2\theta\, \text{Re}\Bigl[\mathcal{B}^{h=0}(x,y)
e^{2\imath(\phi-\chi)}\Bigr]\Biggr\},\\
&\mathcal{A}^{h=0}(x,y) = -\frac12 +
\frac{a_0}{4\pi}\int_0^{2\pi} \frac{\mathrm{d}\alpha}{P},\\
&\mathcal{B}^{h=0}(x,y) = \frac{3a_0}{4\pi}\int_0^{2\pi}
\frac{e^{-2\imath\alpha}\mathrm{d}\alpha}{P},
\end{split}
\end{equation}
Here the exchange length $\ell$ has the standard form \cite{Brown63}:
\begin{equation} \label{eq:exchange-length}
\ell = \sqrt{\frac{\textsf{A}}{4\pi M_s^2}} = a_0\sqrt{\frac{Ja_0^3}{4\pi D}}
\end{equation}
Note that the dipolar induces magnetic anisotropy was considered by
\citet{Levy01} for a pure 2D spin system from a Taylor's series expansion of
the spin field.

The above case \eqref{eq:W-mono} has rather an academic interest. Below in the
paper we consider another limit, when $h\gg a_0$. In that case one can neglect
the energy of the monolayer $W^{h=0}$. The total energy, normalized by the
volume of the magnet, takes a form:
\begin{subequations} \label{eq:W-wol}
\begin{align} \label{eq:W-vol-1}
W &\equiv \frac{\mathcal{E}_{\text{ex}} +
\mathcal{E}_d^h}{M_S^2\mathcal{S}h}
=
W_{\text{ex}} + W_d,\\
\label{eq:W-vol-2} %
W_{\text{ex}} &= \frac{2\pi\ell^2}{\mathcal{S}}\int \mathrm{d}^2x
\Bigl[\left(\vec\nabla \theta\right)^2 +\sin^2\theta\left(\vec\nabla
\phi\right)^2\Bigr],\\
\label{eq:W-vol-3} %
W_d &= \frac{\pi}{\mathcal{S}} \int \mathrm{d}^2 x \Biggl\{
\mathcal{A}(x,y) \left[ 1 - 3 \cos^2\theta\right]\nonumber \\
& + \sin^2\theta\,\text{Re}\Bigl[\mathcal{B}(x,y) e^{2\imath(\phi-\chi)}\Bigr]
\Biggr\}.
\end{align}
\end{subequations}
The effective anisotropy constants can be expressed as follows:
\begin{subequations} \label{eq:A&B}
\begin{align}
\label{eq:A-vol}%
\mathcal{A}(x,y) &\approx \frac{1}{2\pi}\int_0^{2\pi} \mathrm{d}\alpha
\frac{\sqrt{P^2 + h^2}-P}{h} - \frac23,
\\
\label{eq:B-vol} %
\mathcal{B}(x,y) &= \frac{1}{2\pi}\int_0^{2\pi} \mathcal{F}(P,h) e^{-2i\alpha}
\mathrm{d}\alpha,\\
\label{eq:F-vol} %
\mathcal{F}(P,h) &\approx \frac{P - \sqrt{P^2 + h^2}}{h} - 2\ln\frac{\sqrt{P^2
+ h^2}-h}{P}.
\end{align}
\end{subequations}

Let us discuss the magnetization distribution of the nanoparticle on a large
scale. The equilibrium magnetization configuration is mainly determined by the
dipolar interaction, which takes the form of an effective anisotropy
\eqref{eq:W-vol-3}. The coefficient $\mathcal{A}$ determines the uniaxial
anisotropy along the z--axis. For thin nanoparticle this coefficient is always
negative (with $\mathcal{A}\to - 2/3$ when $h \to0$) favoring an easy--plane
magnetization distribution in agreement with the rigorous calculations
\cite{Gioia97}. The coefficient $\mathcal{B}$ is responsible for the in--plane
anisotropy in the XY--plane. Assume that all spins lie in the plane
corresponding to the thin limit case. The preferable magnetization
distribution in the XY--plane is the function $\phi$, minimizing the
expression $\text{Re}\Bigl[\mathcal{B} e^{2\imath(\phi-\chi)}\Bigr]$ in
\eqref{eq:W-wol}. This is
\begin{equation} \label{eq:phi-min}
\phi = \chi + \frac{\pi}{2} - \frac12 \text{Arg}\mathcal{B}.
\end{equation}
The angle \eqref{eq:phi-min} determines the in--plane effective anisotropy
direction observed on a large scale, without exchange interaction and
effective uniaxial anisotropy. The analysis of the $\mathcal{B}$--term shows
that the effective anisotropy favors such an in--plane spin distribution,
always directed tangentially to the border near the sample edge (see Appendix
\ref{sec:appendix-2geom} for the details). This statement agrees with results
for the pure surface anisotropy \cite{Kireev03}. Finer details depend on the
geometry of the particle so we need to distinguish the disk shape from the
square shape.

\subsection{Dispersive part of the dipolar interaction}
\label{sec:disp}

In the continuum description \eqref{eq:S-via-M} the dispersive part of the
dipolar interaction \eqref{eq:Hd-disp} takes the form
\begin{equation*} \label{eq:H-disp-cont}
\begin{split}
& \Delta\mathcal{E}_{\text{d}} = \frac{M_S^2\,a_0^6}{4}\!\!
\int\!\! \mathrm{d}^2x \!\! \int\!\! \mathrm{d}^2x'\!
\Biggl[ A(\vec{r}-\vec{r}')
\Bigl\{\left[\vec{m}(\vec{r})- \vec{m}(\vec{r}')\right]^2\\
&- 3\left[m^z(\vec{r})-m^z(\vec{r}')\right]^2 \Bigr\}
+ B(\vec{r}-\vec{r}') \Bigl\{\left[
m^x(\vec{r})-m^x(\vec{r}')\right]^2\\
& - \left[ m^y(\vec{r})-m^y(\vec{r}')\right]^2 \Bigr\}
+ 2 C(\vec{r}-\vec{r}') \left[ m^x(\vec{r})-m^x(\vec{r}')\right]\\
&\times \left[m^y(\vec{r})-m^y(\vec{r}')\right]\Bigr].
\end{split}
\end{equation*}
By applying the Fourier-transform
\begin{equation} \label{eq:fourier}
\vec{m}(\vec{r})=\frac{1}{(2\pi)^2}\,\int \mathrm{d}^2q\,
\widehat{\vec{m}}(\vec{q})\,e^{\imath\vec{q}\cdot\vec{r}},
\end{equation}
and neglecting finite-size effects, the normalized dispersive part of the
dipole-dipole interaction $\Delta W_d =
{\Delta\mathcal{E}_{\text{d}}}/(M_S^2\mathcal{S}h)$ can be represented in the
form
\begin{equation} \label{eq:disp}
\Delta W_{\text{d}}=\frac{1}{2\pi\mathcal{S}}\int
\mathrm{d}^2q\, \mathfrak{G}(q)\,\Biggl[-|\widehat{m}^z_{\vec{q}}|^2+
\frac{|\vec{q}\cdot\widehat{\vec{m}}_{\vec{q}}|^2}{q^2}\Biggr].
\end{equation}
Here $\vec{q}=\left(q_x,q_y\right)$ is the two-dimensional wave vector,
$\widehat{\vec{m}}(\vec{q})$ is the Fourier-component of the two-dimensional
magnetization $\vec{m}(\vec{r})$, and the function $\mathfrak{G}(q)$ is
defined by the expression
\begin{equation} \label{eq:func}
\mathfrak{G}(q)=\frac{q h-1+e^{-q h}}{qh}.
\end{equation}
Note that Eq.~\eqref{eq:disp} is obtained under assumption that the
ortho-normalization relation
\begin{equation*}
\frac{1}{(2\pi)^2}\int\mathrm{d}^2x\,
\,e^{\imath(\vec{q}-\vec{q}')\cdot\vec{r}}=\delta(\vec{q}-\vec{q}')
\end{equation*}
takes place. Being exact for the infinite domain, this relation is only
approximate for the  finite-size system. For $q h\to0$ the function
\eqref{eq:func} takes the form $\mathfrak{G}(q)\approx q h/2$. Therefore we
expect our approach to yield the correct results for the homogeneous and for
weakly inhomogeneous states. For the general nonhomogeneous spin distribution
the effective anisotropy approach gives only approximate results. In
Sec.~\ref{sec:disk} we verify our effective anisotropy model for disk--shapes
nanoparticles.

\section{Disk--shape nanoparticle}
\label{sec:disk} %

\begin{figure*}
\subfloat[The anisotropy constant $\mathcal{A}(\xi)$ \emph{vs.}
$\xi$.]{\label{fig:A}%
\includegraphics[width=0.49\textwidth]{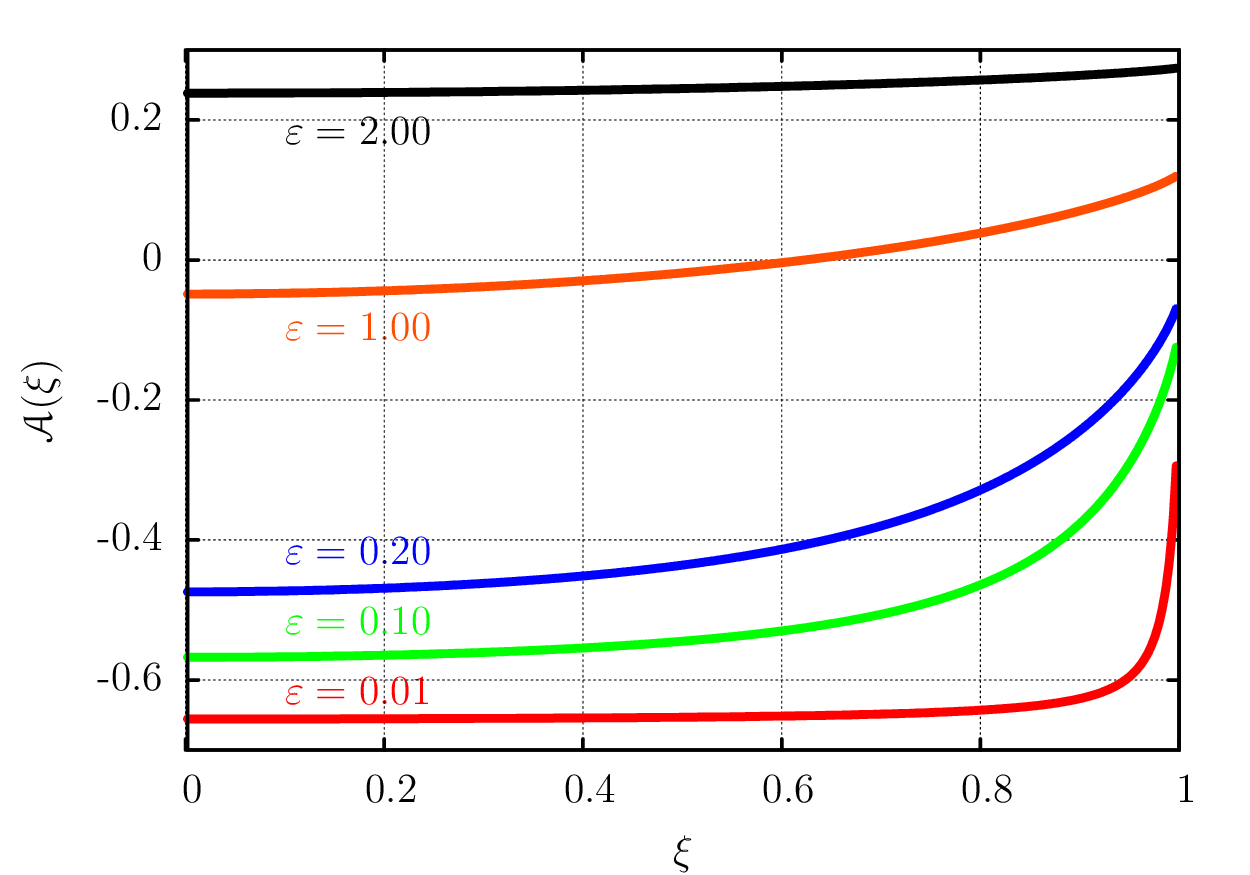}}
\subfloat[The product $\varepsilon \cdot \mathcal{B}(\xi)$ \emph{vs.}
$\xi$. ]{\label{fig:B}%
\includegraphics[width=0.49\textwidth]{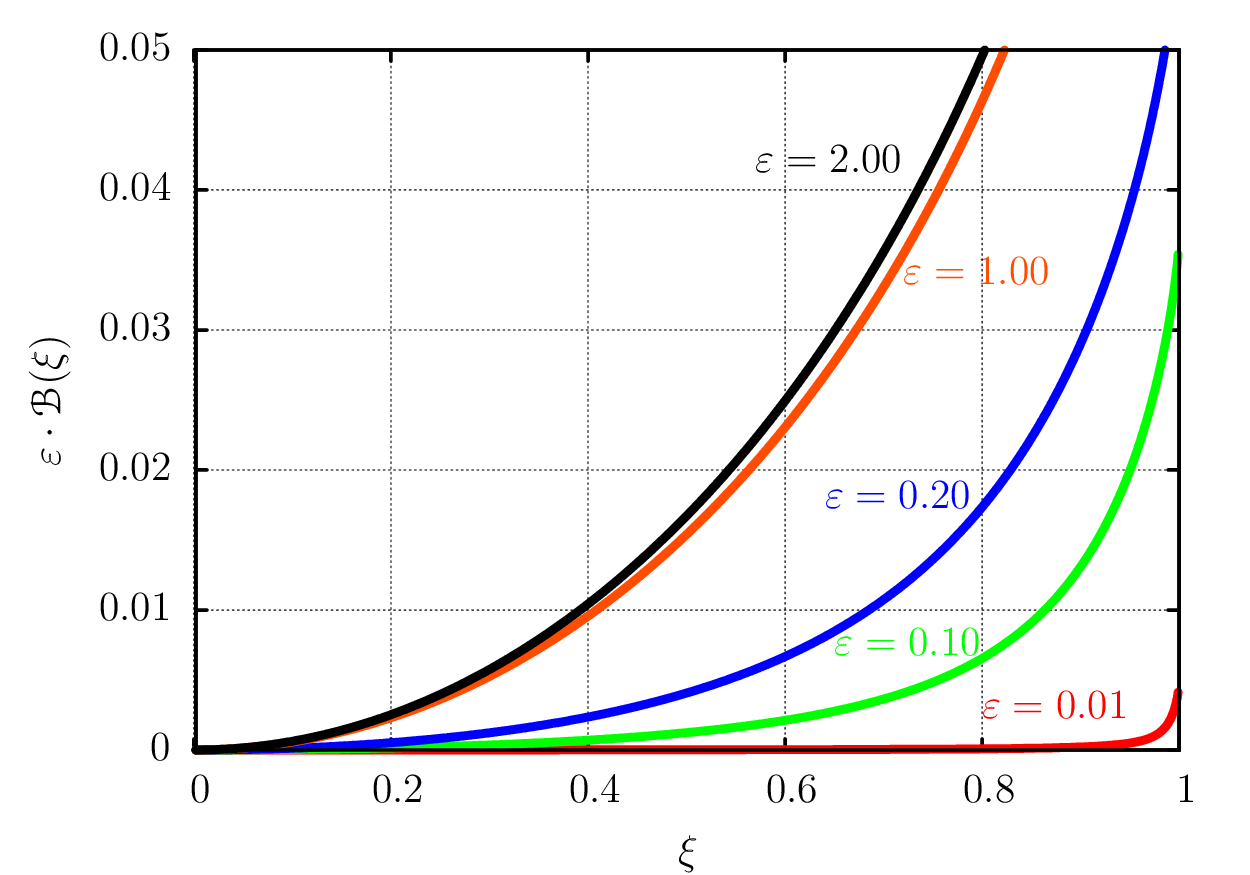}}
\caption{(Color online) Spacial dependence of the effective anisotropy
constants $\mathcal{A}$ [see Eq.~\eqref{eq:A-circular}] and $\mathcal{B}$ [see
Eq.~\eqref{eq:B-circular}].} \label{fig:A-B}
\end{figure*}

Let us consider a cylindric nanoparticle of top surface radius $R$ and
thickness $h$. We introduce $\varepsilon =h/(2R)$ the aspect ratio. Let us
calculate first the effective anisotropy coefficients $\mathcal{A}$ and
$\mathcal{B}$. For the circular system the coefficients $\mathcal{A}$ and
$\mathcal{B}$ depend only on the relative distance $\xi$. We calculated
analytically the coefficients $\mathcal{A}$ and $\mathcal{B}$ (see Appendix
\ref{sec:appendix-continuum}) and these are presented in Fig.~\ref{fig:A-B}
and Eqs.~\eqref{eq:A-circular}, \eqref{eq:B-circular}. First note that when
$\varepsilon \gg1$ both anisotropy constants asymptotically do not depend on
$\xi$: $\mathcal{A}(\xi)\to1/3$ and $\mathcal{B}(\xi)\to0$, see
Fig.~\ref{fig:A-B}. The coefficient of effective uniaxial anisotropy
$\mathcal{A}(\xi)$ slowly depends on $\xi$, namely $\mathcal{A}(0) =
(\sqrt{1+4\varepsilon ^2}-1)/(2\varepsilon)-2/3$ and $\mathcal{A}(1)=1/3$.
When the particle aspect ratio $\varepsilon\lesssim 1$ then
$\mathcal{A}(\xi)<0$, see Fig.~\ref{fig:A} and we have an effective
easy--plane anisotropy. When $\varepsilon\gtrsim 1$, then $\mathcal{A}(\xi)>0$
and we have an effective easy--axis anisotropy. More details are given in
Sec.~\ref{sec:disk-hom}.

In addition to the effective uniaxial anisotropy given by $\mathcal{A}(\xi)$,
we have the essential $\mathcal{B}(\xi)$ term which gives an effective
in--plane anisotropy. For the disk-shaped particle this anisotropy coefficient
is always real, $\arg\mathcal{B}=0$. The value of $\mathcal{B}$ is almost $0$
at origin but its contribution becomes important at the boundary, see
Fig.~\ref{fig:A-B}. We obtain the following asymptotics, valid form small
$\varepsilon$ and $ 1/2 < \xi \lesssim 1$
\begin{align} \label{eq:B-as}
\mathcal{B}(\xi)&\approx \frac{\arctan\Bigl(
\dfrac{\varepsilon}{1-\xi}\Bigr)}{\pi\xi^2}
-\frac{2\varepsilon (3\xi-2)}{3\pi}
\ln\left(\frac{16}{\varepsilon^2+(1-\xi)^2}\right)
\nonumber\\ %
&-\frac{1-\xi}{4\pi\,\varepsilon}
\ln\left(\frac{(1-\xi)^2}{\varepsilon^2+(1-\xi)^2}\right)
\end{align}
(see Appendix \ref{sec:appendix-continuum}). Thus the $\mathcal{B}(\xi)$ term
causes boundary effects and is responsible for the configurational anisotropy.
In the limit $\varepsilon \to0$ (more precisely, when $a_0\ll h\ll R$) the
$\mathcal{B}(\xi)$ term is concentrated near the boundary, corresponding to
the surface anisotropy.

The energy of the nanodisk can be derived from Eq.~\eqref{eq:W-wol}:
\begin{align} \label{eq:W-total-res}
&W = W_{\text{ex}} + W_d,\\
&W_{\text{ex}} = 2\left( \frac{\ell }{R}\right)^2 \int\limits_0^R r
\mathrm{d}r \int\limits_0^{2\pi}\mathrm{d}\chi \Bigl[ (\vec\nabla\theta)^2 +
\sin^2 \theta (\vec\nabla\phi)^2\Bigr],\nonumber\\
&W_d =\! \int\limits_0^{2\pi}\!\! \mathrm{d} \chi \!\!
\int\limits_0^1\!\!\xi\mathrm{d}\xi
\Biggl[\! \mathcal{A}(\xi)\! \left(1\! - \!
3\cos^2\theta \right) + \mathcal{B}(\xi) \sin^2\!\theta
\cos2(\phi\!-\!\chi)\!\Biggr]\!.
\nonumber
\end{align}
In the next subsections we analyze the homogeneous state and the vortex state.

\subsection{Homogeneous state}
\label{sec:disk-hom} %

Let us consider a homogeneous magnetization along the $x$ direction of the
disk--shaped nanodot, so that $\theta=\pi/2$, $\phi=0$. The exchange energy
vanishes. The second term in the dipolar energy \eqref{eq:W-total-res} also
vanishes because of averaging on $\chi$. The total energy $W^x$ is then
\begin{align} \label{eq:Wx}
&W^x = 2\pi \!\!\! \int\limits_0^1\!\!\! \mathcal{A}(\xi) \xi\mathrm{d}\xi =
W_{\text{MS}}^{x}(\varepsilon) -\frac{2\pi}{3}\\
&W_{\text{MS}}^{x}(\varepsilon) = \frac{4}{3\varepsilon}\Bigl\{
-1\! + \! \sqrt{1+\varepsilon^2}\left[\varepsilon^2 \text{K}(m) + \left(1-
\varepsilon^2\right)\text{E}(m) \right] \Bigr\},\nonumber
\end{align}
where $m = (1+\varepsilon^2)^{-1}$, $\text{K}(m)$ and $\text{E}(m)$ are the
complete elliptic integrals of the first and the second kind, respectively
\cite{Abramowitz64}. The constant term $-2\pi/3$ is the isotropic
contribution. The second term $W_{\text{MS}}^{x}$ is the well--known
magnetostatic energy of the homogeneously magnetized disk, first calculated by
\citet{Joseph66}.

If the disk is now homogeneously magnetized along the $z$--axis, then
$\theta=0$. From \eqref{eq:W-total-res} one sees that the corresponding total
energy $W^z = -2W^x$. The transition between these two homogeneous ground
states occurs when $W^x=W^z$. This happens only for $W_x=0$, i.e. for
$W_{\text{MS}}^{x}(\varepsilon_c)=2\pi/3$. This gives a critical value
$\varepsilon_c\approx 0.906$ which agrees with the result by
\citet{Aharoni90}.

\begin{figure}
\includegraphics[width=\columnwidth]{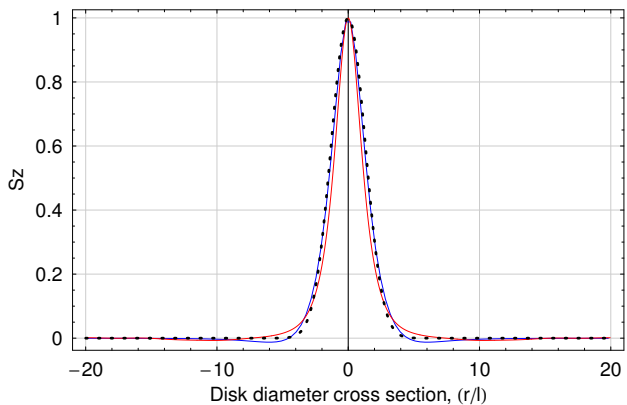}
\caption{(Color online) Comparison of the vortex profiles for the
micromagnetic simulation and the effective anisotropy approximation
for a Py
nanodisk ($2R=212$ nm, $h=16$ nm). The red curve corresponds to the
spin--lattice simulations for the effective anisotropy model with
$\mathcal{H}=\mathcal{H}_{\text{ex}} + \mathcal{H}_d^{loc}$. The
blue curve corresponds to the micromagnetic simulations.
The black dashed curve to the
gaussian ansatz $\cos\theta =
\exp(-r^2/r_v^2)$.}%
\label{fig:diagr-Sz}
\end{figure}

\subsection{Vortex state}
\label{sec:disk-vortex} %

Let us consider a nonhomogeneous state of the disk--shaped particle. In this
state the system has a larger exchange energy compared to the homogeneous
state. This should be compensated by the dipolar term. According to
\eqref{eq:phi-min} the dipolar interaction always favors a spin distribution
of the form
\begin{equation} \label{eq:vortex}
\phi = \chi \pm \frac{\pi}{2},
\end{equation}
where we take into account that the in-plane anisotropy constant $\mathcal{B}$
takes real values only. Such a configuration is called a vortex. In highly
anisotropic magnets there can exist pure planar vortices with $\theta=\pi/2$.
\cite{Wysin94} However we consider here out-of-plane vortices, realized in
"soft" materials typical of nanodisks. The out-of-plane component of the
magnetization has a radial symmetric shape, and it almost does not depend on
$z$ for thin disks, $\theta=\theta(r)$. Now we can calculate the vortex
energy. The vortex solution \eqref{eq:vortex} is characterized by
$\cos2(\phi-\chi)=-1$, providing the minimum of the in--plane component of the
dipolar energy:
\begin{equation} \label{eq:W-dip-vortex}
W_d^{\text{vortex}} = W^x \!-2\pi\!\!\!\int\limits_0^1\!\!
\xi\mathrm{d}\xi \Biggl[3\mathcal{A}(\xi)\cos^2\theta
+ \mathcal{B}(\xi) \sin^2\theta\Biggr]\!.\!
\end{equation}
The exchange energy term
\begin{equation} \label{eq:W-ex-vortex}
\begin{split}
W_{\text{ex}}^{\text{vortex}} = 4\pi \left( \frac{\ell }{R}\right)^2
\int_0^{R}r \mathrm{d}r \left[ {\theta'}^2 + \frac{\sin^2 \theta}{r^2}\right].
\end{split}
\end{equation}
Finally, the vortex energy is
\begin{equation} \label{eq:W-vortex-1}
\begin{split}
W^{\text{vortex}} &=W^x + W_{\text{EP}}^{\text{vortex}} - F(\varepsilon),\\
W_{\text{EP}}^{\text{vortex}} &= 4\pi \left( \frac{\ell }{R}\right)^2 \!\!\!
\int_0^{R} \!\!\! r\mathrm{d}r \Biggl[ {\theta'}^2+ \frac{\sin^2
\theta}{r^2} + \frac{\cos^2\theta}{\ell^2}\Biggr],\\
F(\varepsilon) &= 2\pi\int_0^1\xi\mathrm{d}\xi \Bigl\{
\left[3\mathcal{A}(\xi)+2\right] \cos^2\theta(R\xi)\\
& + \mathcal{B}(\xi) \sin^2\theta(R\xi)\Bigr\}.
\end{split}
\end{equation}
Here $W_{\text{EP}}^{\text{vortex}}$ coincides with the energy of the vortex
in an easy--plane magnet \cite{Ivanov95b},
\begin{equation} \label{eq:Energy-EP}
W_{\text{EP}}^{\text{vortex}} = 2\pi \left( \frac{\ell
}{R}\right)^2\ln\left(\frac{\pi\Lambda R^2}{\ell ^2}\right), \quad
\Lambda=5.27
\end{equation}
and $F(\varepsilon)$ is the configurational anisotropy term. The vortex state
is energetically preferable to the homogeneous state when the configurational
anisotropy term exceeds the energy of the easy--plane vortex $F(\varepsilon) >
W_{\text{EP}}^{\text{vortex}}$. This relation allows to calculate the critical
radius $R_c$ by solving the equation
\begin{equation} \label{eq:critical}
2\pi \left( \frac{\ell }{R}\right)^2\ln\left(\frac{\pi\Lambda R^2} {\ell
^2}\right)= F(\varepsilon).
\end{equation}
To calculate the integral in $F(\varepsilon)$ we use the trial function for
the vortex structure
\begin{equation}\label{eq:ansatz}
m^z\equiv\cos\theta = \exp(-r^2/r_v^2).
\end{equation}
The core width depends on the disk thickness \cite{Kravchuk07}
\begin{equation} \label{eq:rv-vs-h}
r_v(h)\approx \ell\sqrt2 \sqrt[3]{1+ch/\ell}, \qquad c\approx 0.39.
\end{equation}
The relation \eqref{eq:critical} providing the border between the easy-plane
and the out-of-plane vortex states can be analyzed in the limit $\varepsilon
\to0$. Then $F(\varepsilon)\sim (2\pi\varepsilon/3)
\ln\Bigl(\pi/(2\varepsilon) \Bigr)$, hence $R^{(c)}\approx
\ell\sqrt{3/\varepsilon }$. This is in qualitative agreement with previous
results \cite{Usov94,Hoellinger03}.

\begin{figure*}
\subfloat[Spin--lattice simulations for the effective anisotropy model
Hamiltonian \eqref{eq:Hd-loc}.]{\label{fig:square-local}%
\includegraphics[width=0.3\textwidth]{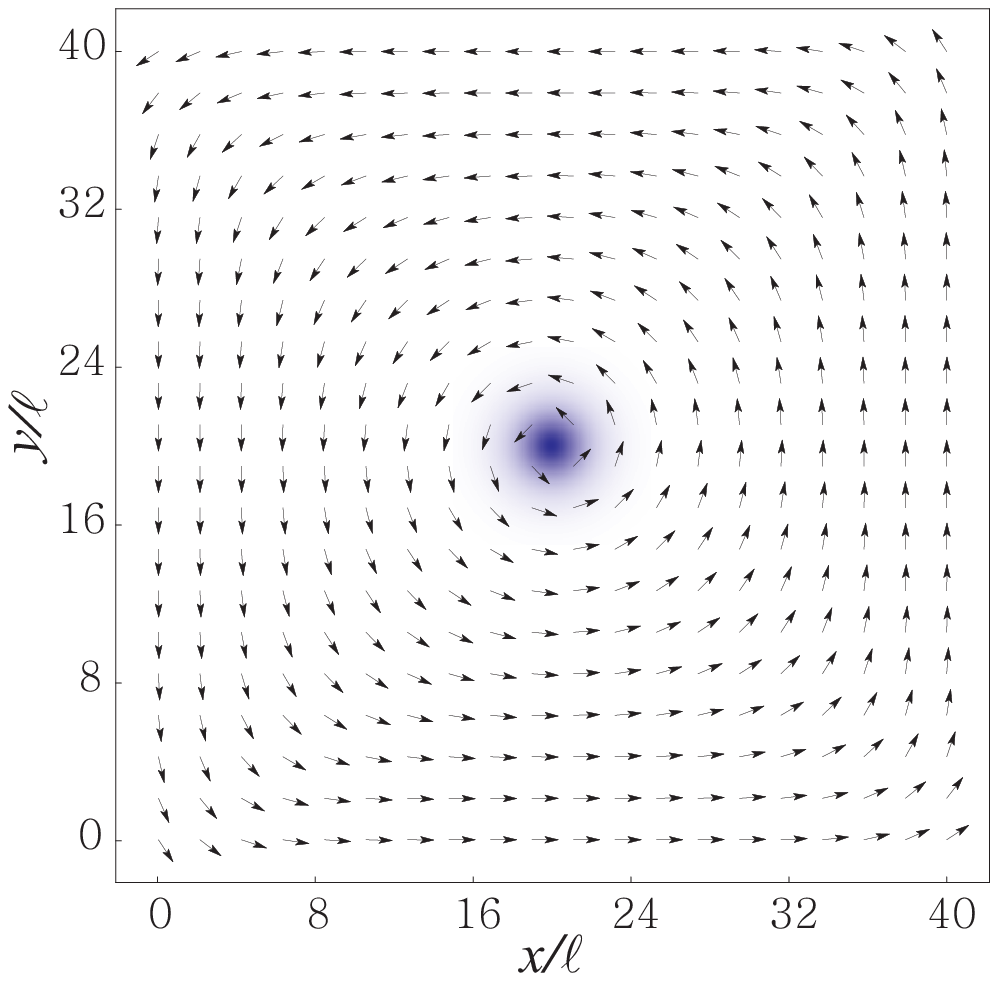}}
\subfloat[Micromagnetic OOMMF simulations data]{\label{fig:squar-OOMMF}%
\includegraphics[width=0.3\textwidth]{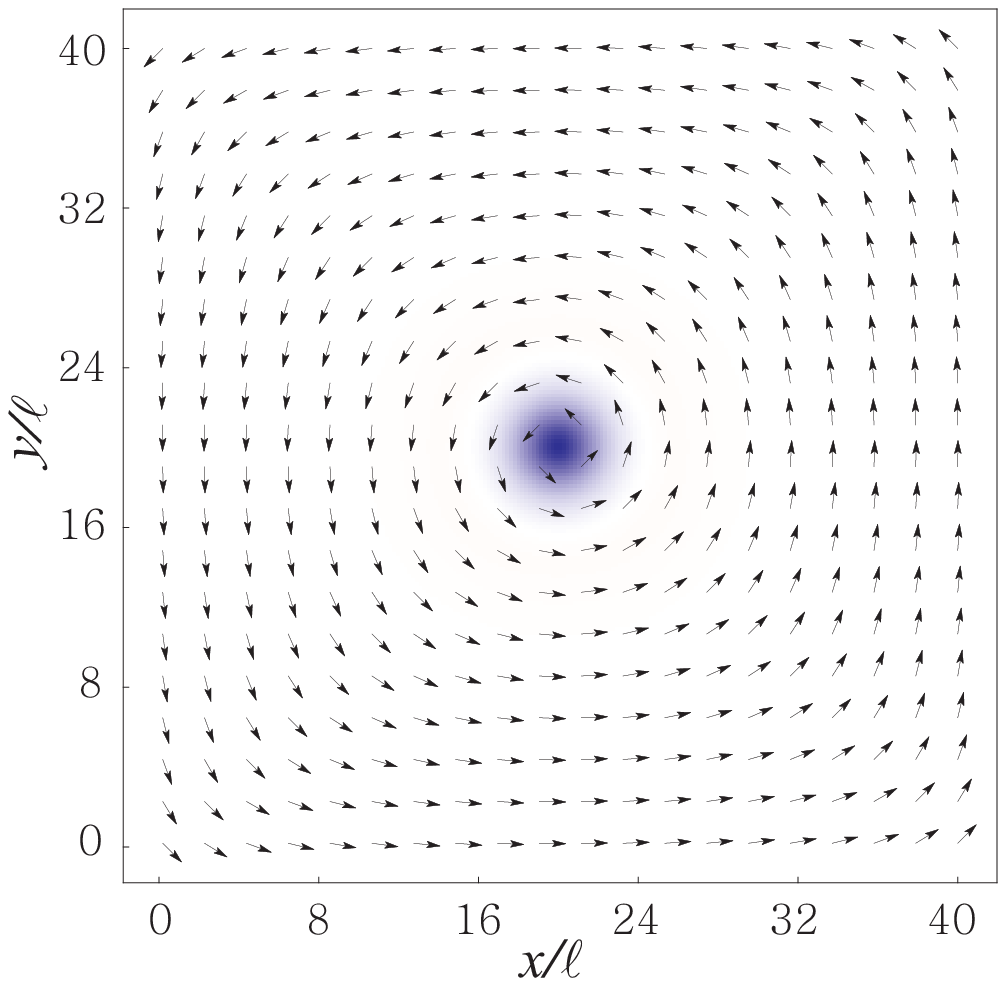}}
\subfloat[Configurational anisotropy lines]{\label{fig:anisLines}%
\includegraphics[width=0.3\textwidth]{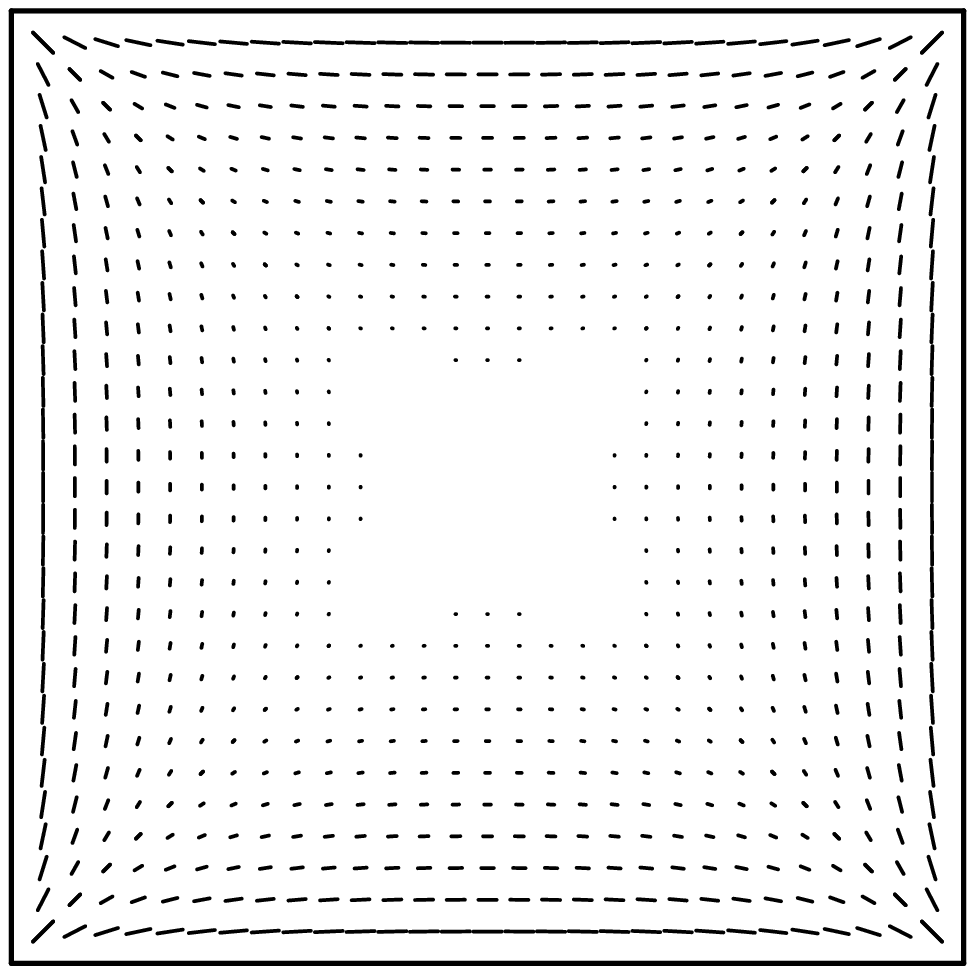}}
\caption{(Color online) Numerical results for the vortex state Py prism (sides
$212\times212$ nm, thickness $h=16$ nm). Figs.~\subref{fig:square-local} and
\subref{fig:squar-OOMMF} represent the spin--field distribution, and
Fig.~\subref{fig:anisLines} the configurational anisotropy lines. These lines
determine the in--plane anisotropy axis orientation, calculated from
Eq.~\eqref{eq:phi-min}; lines lengths correspond to the anisotropy amplitude
in a particular point.}%
\label{fig:diagr-squar}
\end{figure*}

Let us estimate now the contribution of the dispersive part of the dipolar
energy. Taking into account that for the curling state \eqref{eq:phi-min} the
second term in Eq.\eqref{eq:disp} vanishes:
$\vec{q}\cdot\widehat{\vec{m}}_{\vec{q}}\equiv \widehat{\nabla\cdot\vec{m}}=0$
and that the Fourier-component of the out-of-plane component \eqref{eq:ansatz}
has the form $\widehat{m}^z=\pi\,r^2_v\,e^{-q^2r^2_v}$, from
Eq.\eqref{eq:disp} we get
\begin{equation}\label{eq:disp-contr}
\Delta W_{\text{d}}\approx
\begin{cases}
\dfrac{\sqrt{\pi}}{8}\,\varepsilon\,\dfrac{r_v}{R} & \text{for $r_v\gg h$}, \\
\dfrac{1}{2}\,\dfrac{r_v^2}{R^2} & \text{for $r_v\ll h$}.
\end{cases}
\end{equation}
Comparing Eq.~\eqref{eq:disp-contr} with Eqs.~\eqref{eq:W-vortex-1} and
\eqref{eq:Energy-EP}, and taking into account \eqref{eq:rv-vs-h}, one can
conclude that in the limit $\varepsilon \to0$ the dispersive part of the
dipolar interaction does not change significantly the vortex stability
criterion. More precisely, our effective anisotropy approximation works
correctly not only for $\varepsilon \to0$ but also for $R\gg h,~R\gg \ell$.
Our numerical results show that it gives the vortex state as an energy minimum
for disk diameters $2R\gtrsim 30\ell$.

\section{Numerical simulations}
\label{sec:simulations} %

To check our effective anisotropy approximation, we performed numerical
simulations. We used the publicly available three--dimensional OOMMF
micromagnetic simulator code \cite{OOMMF}. In all micromagnetic simulations we
used the following material parameters for Py: $A=1.3\times10^{-6}$ erg/cm
(using SI units $A^{\text{SI}}=1.3\times 10^{-11}$ J/m), $M_s=8.6\times10^{2}$
G ($M_s^{\text{SI}}=8.6\times 10^5$ A/m), the damping coefficient $\eta =
0.006$, and the anisotropy has been neglected. This corresponds to an exchange
length $\ell = \sqrt{A/4\pi M_s^2}\approx 5.3$nm ($\ell^{\text{SI}} =
\sqrt{A/\mu_0 M_s^2}$). The mesh cells were cubic (2 nm).

We also test our effective anisotropy approach by the \emph{original discrete
spin--lattice simulator}. The spin dynamics is described by the discrete
version of the Landau--Lifshitz equations with Gilbert damping
\begin{equation} \label{eq:LL-discrete}
\frac{\mathrm{d} \vec{S}_{\vec{n}} }{\mathrm{d} t} = -
\left[\vec{S}_{\vec{n}}\times \frac{\partial \mathcal{H} }{\partial
\vec{S}_{\vec{n}}}\right] - \frac{\eta}{S} \left[\vec{S}_{\vec{n}} \times
\frac{\mathrm{d} \vec{S}_{\vec{n}} }{\mathrm{d}t}\right],
\end{equation}
which we consider on a 2D square lattice of size $(2R)^2$. We have assumed a
plane--parallel spin distribution homogeneous along the z--direction. Each
lattice is bounded by a circle of radius $R$ on which the spins are free
corresponding to a Neuman boundary condition in the continuum limit. We
integrate the discrete Landau--Lifshitz equations \eqref{eq:LL-discrete} with
the Hamiltonian $\mathcal{H}=\mathcal{H}_{\text{ex}} + \mathcal{H}_d$ given by
\eqref{eq:H-ex} and \eqref{eq:H-dipolar-via-ABC}, using a 4th--order
Runge--Kutta scheme with time step $0.01/N_z$. These spin--lattice simulations
were done to validate our analytical calculations for the effective anisotropy
model. Throughout this work we compared the results of the spin--lattice
simulations with $\mathcal{H}=\mathcal{H}_{\text{ex}} + \mathcal{H}_d$ with
the results of micromagnetic simulations. We never found any noticeable
difference. We present the results for a disk--shaped and a prism--shaped
nanoparticle because these two geometries are the most common ones in
experiments.

\begin{figure*}
\subfloat[Distance from the vortex center $r=10\ell$.]{\label{fig:squareR1}%
\includegraphics[width=0.3\textwidth]{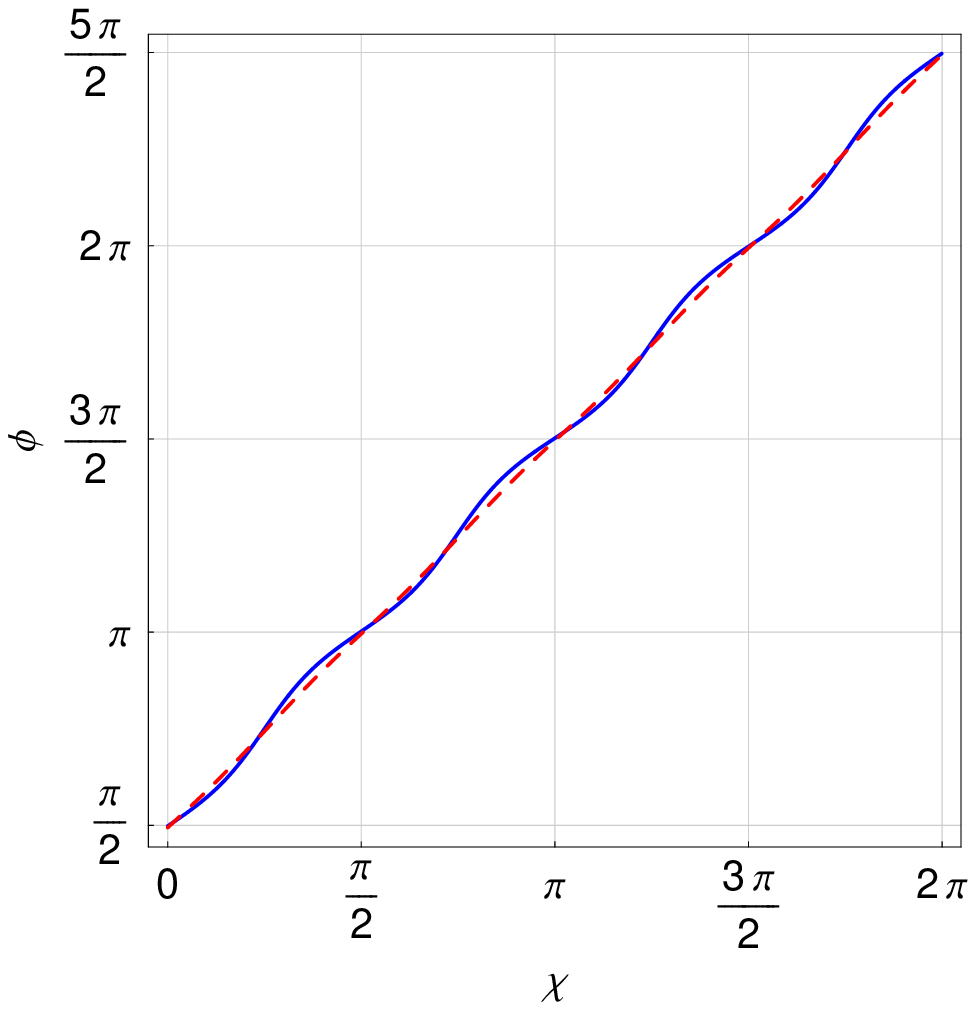}}
\subfloat[Distance from the vortex center $r=20\ell$.]{\label{fig:squareR6}%
\includegraphics[width=0.3\textwidth]{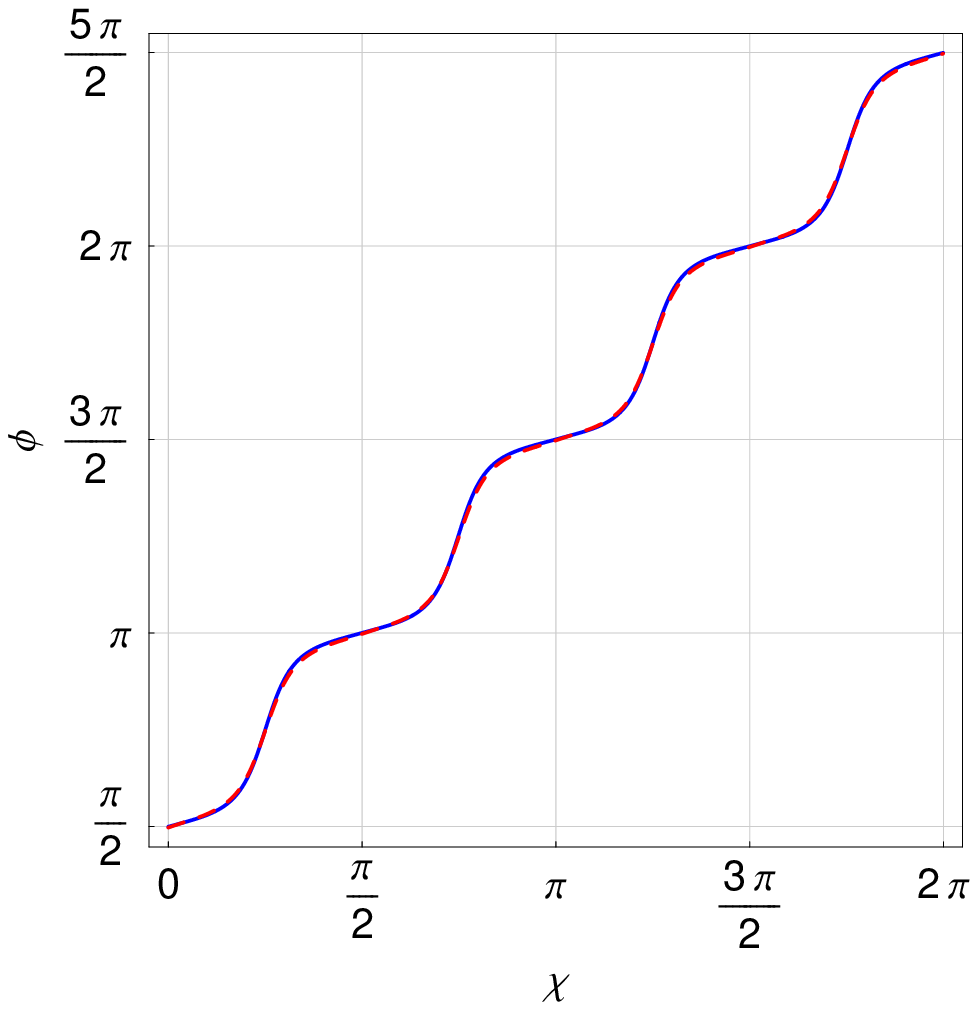}}
\caption{(Color online) The in--plane spin angle $\phi$ as a function of the
polar angle for the vortex state in a prism of Py of sides $212\times212$ nm
and thickness $h=16$ nm. The red dashed curves correspond to the effective
anisotropy approximation, the blue solid curves
--- to the micromagnetic simulations data.} \label{fig:squareR}
\end{figure*}

\subsection{Disk--shape nanoparticle}
\label{sec:disk-simulations}%

Our effective anisotropy approximation provides the exact solution for all
homogeneous states for a nanodisk. Therefore we do not need to justify it for
the homogeneous states. We consider here the vortex state. As we analyzed
before, the model can provide the preferable vortex state for disk diameters
$2R>30\ell$, which is in an agreement of the model usage criterium
\eqref{eq:effective-correct}. We compare the magnetization distribution in the
vortex for our effective anisotropy model and for the the micromagnetic
simulations. Since the in--plane vortex structure is characterized by the same
distribution $\phi=\chi\pm\pi/2$ for both methods, we are interested in the
out--of--plane vortex profiles. We performed such a comparison for a disk of
size $2R/\ell=40$ and $h/\ell=3$, which satisfy the criterium
\eqref{eq:effective-correct}. The results are presented on
Fig.~\ref{fig:diagr-Sz}. One can see that the vortex shape from the effective
anisotropy model agrees with the one obtained from the micromagnetic
simulations within $0.11$ in absolute error.

\subsection{Prism--shape nanoparticle}
\label{sec:prism} %

Now we check the validity of the effective anisotropy approximation for the
prism--shaped nanoparticle. We chose this shape because there are numerous
experiment with a square geometry, see for a review
Ref.~\onlinecite{Hubert98}. We performed the two types of simulations for a
square shaped nanoparticle, see Figs.~\ref{fig:square-local} and
\ref{fig:squar-OOMMF}. The two equilibrium magnetization distributions,
obtained for the micromagnetic model and the spin-lattice simulation agree
with a very high precision.

As discussed above the large scale distribution of the magnetization is
described by Eq.~\eqref{eq:phi-min}. Calculating numerically the coefficient
$\mathcal{B}$ (see Appendix \ref{sec:appendix-2geom} for details), we found
the distribution of the configurational anisotropy lines for the square
geometry. This is shown in Fig.~\ref{fig:anisLines}. The comparison of
Figs.~\ref{fig:diagr-squar} shows that the effective anisotropy lines
corresponds to the magnetization direction in the main part of the system.
Note that the effective anisotropy approach fails near the corners: the sharp
field distribution near the prism vertices (Fig.~\ref{fig:anisLines}) is not
energetically preferable when the exchange contribution is taken into account.

We can also check the validity of the effective anisotropy approach for the
complicated ``vortex'' structure in the square geometry, by comparing the
distribution of the in-plane spin angle $\phi$ to the one given by the
micromagnetic simulations. This is done in Fig~\ref{fig:squareR}. The figure
shows that the two different approaches agree very well. The
$\phi(\chi)$--dependencies coincide within $0.11$ in absolute error for
$r=10\ell$ and within $0.04$ for $r=20\ell$.

\section{Discussion}
\label{sec:discussion} %

To summarize, assuming that magnetization is independant of the thickness
variable $z$, we have reduced the magnetic energy of a thin nanodot to a local
2D inhomogeneous anisotropy. The first term $\mathcal{A}$ determines the
uniaxial anisotropy along the $z$--axis. The second term $\mathcal{B}$ gives
the anisotropy in the $XY$--plane.

For thin nanoparticles $\varepsilon \lesssim 1$ the term
$\mathcal{A}\approx\text{const}<0$, gives an effective easy--plane anisotropy.
This generalizes the rigorous results obtained for infinitesimally thin films
\cite{Gioia97}. The function $\mathcal{B}(x,y)$ is localized near the edge of
the particle so that spins will be tangent to the boundary. This confirms the
notion of a surface edge anisotropy
\cite{Kireev03,Tchernyshyov05}. \\
When the nanoparticle is thick $\varepsilon \gtrsim 1$, the anisotropy
constant $\mathcal{A}>0$, is again almost constant and the spins will tend to
follow the $z$ axis (easy--axis anisotropy). The in-plane anisotropy
$\mathcal{B}$ depends on the thickness, see Fig.~\ref{fig:A}. The special
distribution of $\mathcal{B}(x,y)$ is responsible for the volume contribution
of the dipolar energy.

The above effective anisotropy approach: (i) shows the nature of the effective
easy-plane anisotropy and the surface anisotropy, (ii) generalizes the surface
anisotropy for the finite thickness, and (iii) gives a unified approach to
study dipolar effects in pure 2D systems and 3D magnets of finite thickness.

It is instructive to make a link between our approach and the rigorous results
which were obtained in
Refs.~\onlinecite{Desimone95,Desimone02,Gioia97,Moser04,Kohn05,Kohn05a,Kurzke06}.
Our equations \eqref{eq:W-vortex-1}, \eqref{eq:critical} show that for the
vortex ground state to exist, it is crucial to have both types of anisotropy:
out-of-plane anisotropy  and in-plane one. It is shown by \citet{Kohn05a} that
the energy of a thin magnetic film with an accuracy up to $\varepsilon^2$ can
be presented as the sum
\begin{equation} \label{eq:Kohn}
\begin{split}
&E=E_{\text{exch}}+E_{\text{bdry}}+E_{\text{trans}}\\
&=\ell^2 \varepsilon \!\!\int_{\omega}\!\! |\nabla
\vec{m}|^2+\frac{\varepsilon^2|\ln\varepsilon|}{2\pi}
\!\!\int_{\partial\omega} \!\! \left(\vec{m}\cdot\vec{n}\right)^2+
\varepsilon \!\! \int_{\omega} \!\! \left(m^z\right)^2.
\end{split}
\end{equation}
Considering the limit $\varepsilon\to0$ and
$\ell^2/(\varepsilon|\ln\varepsilon|)=\text{const}$ we see from
Eq.~\eqref{eq:Kohn} that formally the last term is dominating and its
contribution has to be accounted as a constraint $m^z=0$, see
Ref.~\onlinecite{Kohn05a}. This constraint prevents the existence of the
vortex ground state of the nanodot because the energy of the vortex in the
continuum limit is infinite due to divergence at $r\to0$. However this
divergence is removed by the out-of-plane component of vortex which is
described by a localized function with radius of localization $r_v\sim \ell$
[see Eq.~\eqref{eq:rv-vs-h}]. This means that the last term $E_{\text{trans}}$
in \eqref{eq:Kohn} scales like the exchange term $E_{\text{exch}}$. In this
limit all three terms of \eqref{eq:Kohn} are of the same order and provide the
existence of the vortex ground state.

This reduction of the nonlocal dipolar interaction to a local form is a first
step towards an analytical study of nanomagnetism. We developed a method of
effective anisotropy and illustrated it on a few examples. We plan to apply
this method to the dynamics of vortices in nanomagnets.

\acknowledgments

Yu.G., V.P.K. and D.D.S. thank the University of Bayreuth, where part of this
work was performed, for kind hospitality and acknowledge the support from
Deutsches Zentrum f{\"u}r Luft- und Raumfart e.V., Internationales B{\"u}ro
des BMBF in the frame of a bilateral scientific cooperation between Ukraine
and Germany, project No.~UKR~05/055. J.G.C., Yu.G. and D.D.S. acknowledge
support from a Ukrainian--French Dnipro grant (No.~82/240293). Yu. G. thanks
the University of Cergy-Pontoise for an invited professorship during which
this work was completed. D.D.S. acknowledges the support from the Alexander
von Humboldt--Foundation. V.P.K. acknowledges the support from the BAYHOST
project. J.G.C. thanks the Centre de Ressources Informatiques de
Haute-Normandie where part of the computations were carried out.

\appendix

\begin{widetext}

\section{Discrete dipolar energy calculations}
\label{sec:appendix-discrete}

Let us consider the dipolar interaction term $\mathcal{H}_{\text{d}}$. Using
the notations
\begin{equation} \label{eq:notations} %
\begin{split}
\frac{x_{\vec n \vec m}}{a_0} = n_x - m_x,\; \frac{y_{\vec n \vec m}}{a_0} =
n_y - m_y,\; \frac{z_{\vec n \vec m}}{a_0} = n_z - m_z,\; \rho_{{\vec{\nu}}
{\vec{\mu}}} = \sqrt{x_{\vec n\vec m}^2 + y_{\vec n \vec m}^2}, \; r_{\vec
n\vec m} = \sqrt{\rho_{{\vec{\nu}} {\vec{\mu}}}^2 + z_{\vec n\vec m}^2},
\end{split}
\end{equation}
one can rewrite this energy as follows:
\begin{equation} \label{eq:H-dipolar-1}
\begin{split}
\mathcal{H}_{\text{d}} &= \frac{D}{2}\!\!
\sum_{\substack{\vec{n},\vec{m}\\r_{\vec{n}\vec{m}}\neq0}}\! \Biggl\{
\frac{\left(\vec{S}_{\vec{n}}\cdot
\vec{S}_{\vec{m}}\right)}{r_{\vec{n}\vec{m}}^3} -
\frac{3S_{\vec{n}}^zS_{\vec{m}}^z z_{\vec{n}\vec{m}}^2}{r_{\vec{n}\vec{m}}^5}
- \frac{6}{r_{\vec{n}\vec{m}}^5}
S_{\vec{n}}^z z_{\vec{n}\vec{m}}\left(S_{\vec{m}}^x x_{\vec{n}\vec{m}}+ S_{\vec{m}}^y y_{\vec{n}\vec{m}}\right)\\
& -\frac{3}{r_{\vec{n}\vec{m}}^5}\left( S_{\vec{n}}^x x_{\vec{n}\vec{m}} +
S_{\vec{n}}^y y_{\vec{n}\vec{m}} \right) \left( S_{\vec{m}}^x
x_{\vec{n}\vec{m}} + S_{\vec{m}}^y y_{\vec{n}\vec{m}}
\right)\Biggr\} = D\!\! \sum_{\substack{{\vec{\nu}},{\vec{\mu}}\\
\rho_{{\vec{\nu}}{\vec{\mu}}}\neq0}}\! \Biggl\{
S_{{\vec{\nu}}}^zS_{{\vec{\mu}}}^z K_z(\rho_{{\vec{\nu}}{\vec{\mu}}}) + \left(
S_{{\vec{\nu}}}^xS_{{\vec{\mu}}}^x
+ S_{{\vec{\nu}}}^yS_{{\vec{\mu}}}^y \right) K_1(\rho_{{\vec{\nu}}{\vec{\mu}}})\\
&- \left( S_{{\vec{\nu}}}^x x_{{\vec{\nu}}{\vec{\mu}}} + S_{{\vec{\nu}}}^y
y_{{\vec{\nu}}{\vec{\mu}}} \right) \left( S_{{\vec{\mu}}}^x
x_{{\vec{\nu}}{\vec{\mu}}} + S_{{\vec{\mu}}}^y y_{{\vec{\nu}}{\vec{\mu}}}
\right)K_2(\rho_{{\vec{\nu}}{\vec{\mu}}})\Biggr\}.
\end{split}
\end{equation}
Here we used the obvious relations
$x_{\vec{n}\vec{m}}=x_{{\vec{\nu}}{\vec{\mu}}}$,
$y_{\vec{n}\vec{m}}=y_{{\vec{\nu}}{\vec{\mu}}}$ and the basic assumption that
the magnetization does not depend on the z-coordinate:
$\vec{S}_{\vec{n}}=\vec{S}_{{\vec{\nu}}}$,
$\vec{S}_{\vec{m}}=\vec{S}_{{\vec{\mu}}}$. This allows us to reduce the
summation to the 2D lattice. The kernels $K_1$, $K_2$ and $K_z$ contain
information about the original 3D structure of our system,
\begin{equation} \label{eq:K-1-2-3}
\begin{split}
K_1(s) &= \frac12 \sum_{n_z, m_z}\frac{1}{\left(s^2 + z_{\vec n\vec
m}^2\right)^{3/2}},\quad K_2(s) = \frac32 \sum_{n_z, m_z}\frac{1}{\left(s^2 +
z_{\vec n\vec m}^2\right)^{5/2}},\quad K_z(s) = \frac12 \sum_{n_z,
m_z}\frac{s^2 - 2z_{\vec n\vec m}^2}{\left(s^2 + z_{\vec n\vec
m}^2\right)^{5/2}}.
\end{split}
\end{equation}
Taking into account that
\begin{equation*}
\begin{split}
& S_{{\vec{\nu}}}^x S_{{\vec{\mu}}}^x x_{{\vec{\nu}}{\vec{\mu}}}^2 +
S_{{\vec{\nu}}}^y S_{{\vec{\mu}}}^y y_{{\vec{\nu}}{\vec{\mu}}}^2= \tfrac12
\rho_{{\vec{\nu}}{\vec{\mu}}}^2 \left(S_{{\vec{\nu}}}^x S_{{\vec{\mu}}}^x +
S_{{\vec{\nu}}}^y S_{{\vec{\mu}}}^y\right) + \tfrac12\left(
x_{{\vec{\nu}}{\vec{\mu}}}^2 - y_{{\vec{\nu}}{\vec{\mu}}}^2 \right)
\left(S_{{\vec{\nu}}}^x S_{{\vec{\mu}}}^x - S_{{\vec{\nu}}}^y
S_{{\vec{\mu}}}^y\right),
\end{split}
\end{equation*}
one can present the dipolar energy in more symmetrical way:
\begin{equation} \label{eq:H-dipolar-K2&Kz}
\begin{split}
\mathcal{H}_{\text{d}} = -\frac{D}{2}\!\!
\sum_{\substack{{\vec{\nu}},{\vec{\mu}}\\
\rho_{{\vec{\nu}}{\vec{\mu}}}\neq0}}\! \Biggl\{ &
K_z(\rho_{{\vec{\nu}}{\vec{\mu}}})\left(\vec{S}_{{\vec{\nu}}}\cdot
\vec{S}_{{\vec{\mu}}} - 3 S_{{\vec{\nu}}}^zS_{{\vec{\mu}}}^z\right)+
K_2(\rho_{{\vec{\nu}}{\vec{\mu}}})\left(x_{{\vec{\nu}}{\vec{\mu}}}^2 -
y_{{\vec{\nu}}{\vec{\mu}}}^2 \right)\left( S_{{\vec{\nu}}}^x S_{{\vec{\mu}}}^x
- S_{{\vec{\nu}}}^y S_{{\vec{\mu}}}^y
\right) \\
&+ 2 K_2(\rho_{{\vec{\nu}}{\vec{\mu}}}) x_{{\vec{\nu}}{\vec{\mu}}}
y_{{\vec{\nu}}{\vec{\mu}}}\left( S_{{\vec{\nu}}}^x S_{{\vec{\mu}}}^y +
S_{{\vec{\nu}}}^y S_{{\vec{\mu}}}^x \right)\Biggr\}.
\end{split}
\end{equation}
The total Hamiltonian is the sum of two terms \eqref{eq:H-ex} and
\eqref{eq:H-dipolar-K2&Kz}.

Here we show that the main effect of the nonlocal dipolar interaction is an
effective nonhomogeneous anisotropy. Using equality
\begin{equation*} \label{eq:equality}
\begin{split}
&\sum_{\vec{n},\vec{m}} C_{\vec{m}\vec{n}}S_{\vec{n}} S_{\vec{m}} =
\sum_{\vec{n}} \mathcal{C}_{\vec{n}}S_{\vec{n}}^2 - \frac12
\sum_{{\vec{n}},{\vec{m}}} C_{{\vec{n}}{\vec{m}}}\left( S_{\vec{n}} -
S_{\vec{m}} \right)^2, \quad \mathcal{C}_{\vec{n}} = \sum_{{\vec{m}}}
C_{{\vec{n}}{\vec{m}}},
\end{split}
\end{equation*}
where $C_{{\vec{n}}{\vec{m}}} = C_{{\vec{m}}{\vec{n}}}$, one can split the
dipolar Hamiltonian \eqref{eq:H-dipolar-K2&Kz} into a local contribution and a
nonlocal correction
\begin{align}
\tag{\ref{eq:Hd-loc}}
\label{eq:Hd-loc-again} %
\mathcal{H}_{\text{d}} &= \mathcal{H}_{\text{d}}^{\text{loc}} +
\Delta\mathcal{H}_{\text{d}},\quad \mathcal{H}_{\text{d}}^{\text{loc}} =
-\frac{D}{2}\!\! \sum_{{\vec{\nu}}}\! \Biggl\{
\bar{A}_{{\vec{\nu}}}\Bigl[\left(\vec{S}_{{\vec{\nu}}}\right)^2 - 3\left(
S_{{\vec{\nu}}}^z\right)^2\Bigr] + \bar{B}_{{\vec{\nu}}}
\Bigl[\left(S_{{\vec{\nu}}}^x\right)^2 - \left(S_{{\vec{\nu}}}^y\right)^2
\Bigr] + 2\bar{C}_{{\vec{\nu}}} S_{{\vec{\nu}}}^x
S_{{\vec{\nu}}}^y\Biggr\},\\
\label{eq:H-dipolar-via-local} %
\Delta \mathcal{H}_{\text{d}} &= \frac{D}{4}\!\!
\sum_{\substack{{\vec{\nu}},{\vec{\mu}}\\
\rho_{{\vec{\nu}}{\vec{\mu}}}\neq0}}\! \Biggl\{
K_z(\rho_{{\vec{\nu}}{\vec{\mu}}})\Bigl[\left(\vec{S}_{{\vec{\nu}}}-
\vec{S}_{{\vec{\mu}}}\right)^2 - 3 \left(S_{{\vec{\nu}}}^z -
S_{{\vec{\mu}}}^z\right)^2\Bigr]+
K_2(\rho_{{\vec{\nu}}{\vec{\mu}}})\left(x_{{\vec{\nu}}{\vec{\mu}}}^2 -
y_{{\vec{\nu}}{\vec{\mu}}}^2 \right)\Bigl[ \left( S_{{\vec{\nu}}}^x -
S_{{\vec{\mu}}}^x\right)^2 - \left(S_{{\vec{\nu}}}^y - S_{{\vec{\mu}}}^y
\right)^2\Bigr] \nonumber\\
&+ 4 K_2(\rho_{{\vec{\nu}}{\vec{\mu}}}) x_{{\vec{\nu}}{\vec{\mu}}}
y_{{\vec{\nu}}{\vec{\mu}}}\Bigl[\left( S_{{\vec{\nu}}}^x -
S_{{\vec{\mu}}}^x\right) \left(S_{{\vec{\nu}}}^y - S_{{\vec{\mu}}}^y
\right)\Bigr]\Biggr\}.
\end{align}

\section{Continuum limit of the local dipolar energy}
\label{sec:appendix-continuum}%

Here we present the continuum limit of the discrete dipolar Hamiltonian
\eqref{eq:Hd-loc} corresponding to the dipolar energy
\begin{equation} \label{eq:Energy-dipolar} %
\begin{split}
\mathcal{E}_{\text{d}}^{\text{loc}} &= -\frac{a_0^6 M_S^2}{2}\!\!
\sum_{{\vec{\nu}}}\! \Biggl\{ \bar{A}_{{\vec{\nu}}}\Bigl[1 -
3\left( m_{{\vec{\nu}}}^z\right)^2\Bigr] + \bar{B}_{{\vec{\nu}}}
\Bigl[\left(m_{{\vec{\nu}}}^x\right)^2 -
\left(m_{{\vec{\nu}}}^y\right)^2 \Bigr] + 2\bar{C}_{{\vec{\nu}}}
m_{{\vec{\nu}}}^x m_{{\vec{\nu}}}^y\Biggr\},
\end{split}
\end{equation}
where $\vec{m}_{\vec{\nu}} = \frac{g\mu_B}{a_0^3 M_s}\vec{S}_{\vec{\nu}}$.
Hence the continuous magnetization vector $\vec{m}$ according to
Eq.~\eqref{eq:S-via-M} takes the form $\vec{m}(\vec{r}) =
\sum_{\vec{\nu}}\vec{m}_{\vec{\nu}} \delta \left(\vec{r} - \vec{r}_{\vec{\nu}}
\right)$. Here $\bar{A}_{{\vec{\nu}}}$, $\bar{B}_{{\vec{\nu}}}$ and
$\bar{C}_{{\vec{\nu}}}$ are determined as follows
\begin{subequations} \label{eq:A-B-C-details}
\begin{align} \label{eq:A-details}
\bar{A}_{{\vec{\nu}}} &=
\sum_{\substack{{\vec{\mu}}\\r_{\vec{n}\vec{m}}\neq0}}
K_z(\rho_{{\vec{\nu}}{\vec{\mu}}}) = \frac12
\sum_{\substack{{\vec{\mu}}\\r_{\vec{n}\vec{m}}\neq0}}
\sum_{n_z,m_z} \frac{\rho_{{\vec{\nu}}{\vec{\mu}}}^2 -
2z_{\vec{n}\vec{m}}^2}{\left(\rho_{{\vec{\nu}}{\vec{\mu}}}^2
+ z_{\vec{n}\vec{m}}^2\right)^{5/2}},\\
\label{eq:B-details} %
\bar{B}_{{\vec{\nu}}} &=
\sum_{\substack{{\vec{\mu}}\\r_{\vec{n}\vec{m}}\neq0}}
K_2(\rho_{{\vec{\nu}}{\vec{\mu}}})
\left(x_{{\vec{\nu}}{\vec{\mu}}}^2 - y_{{\vec{\nu}}{\vec{\mu}}}^2
\right)= \frac32
\sum_{\substack{{\vec{\mu}}\\r_{\vec{n}\vec{m}}\neq0}}
\sum_{n_z,m_z} \frac{x_{{\vec{\nu}}{\vec{\mu}}}^2 -
y_{{\vec{\nu}}{\vec{\mu}}}^2}{\left(\rho_{{\vec{\nu}}{\vec{\mu}}}^2
+ z_{\vec{n}\vec{m}}^2\right)^{5/2}},\\
\label{eq:C-details} %
\bar{C}_{{\vec{\nu}}} &=
\sum_{\substack{{\vec{\mu}}\\r_{\vec{n}\vec{m}}\neq0}}
K_2(\rho_{{\vec{\nu}}{\vec{\mu}}}) 2x_{{\vec{\nu}}{\vec{\mu}}}
y_{{\vec{\nu}}{\vec{\mu}}} = \frac32
\sum_{\substack{{\vec{\mu}}\\r_{\vec{n}\vec{m}}\neq0}}
\sum_{n_z,m_z} \frac{2x_{{\vec{\nu}}{\vec{\mu}}}
y_{{\vec{\nu}}{\vec{\mu}}} }{\left(\rho_{{\vec{\nu}}{\vec{\mu}}}^2 +
z_{\vec{n}\vec{m}}^2\right)^{5/2}}.
\end{align}
\end{subequations}
The continuum version of the effective anisotropy constants
\eqref{eq:A-B-C-details} can be found using a relation
\begin{equation} \label{eq:int-relation}
\begin{split}
\sum_{n_z=0}^{N_z} \sum_{m_z=0}^{N_z} F(|z_{\vec{nm}}|) &\approx
\frac{1}{a_0^2} \int_0^h \mathrm{d} z \int_0^h \mathrm{d} z'
F(|z-z'|) + \frac{1}{a_0} \int_0^h \mathrm{d} z \bigl[F(|z|) +
F(|h-z|)\bigr] + \frac12
\left[ F(0) + F(|h|)\right] \\
& = \frac{2}{a_0^2} \int_0^h \mathrm{d} z F(|z|) \bigl[h-z+a_0\bigr]
+ \frac12 \left[ F(0) + F(|h|)\right], \qquad h=N_z a_0\geq0.
\end{split}
\end{equation}

Let us start with the calculation of the coefficient $\bar{A}_{{\vec{\nu}}}$
from Eq.~\eqref{eq:A-details}:
\begin{align} \label{eq:A-calculation}
&\mathcal{A}(x,y) \equiv - \frac{a_0^4}{2\pi
h}\bar{A}_{{\vec{\nu}}} = \frac{1}{h}\left(\mathcal{A}_1 +
\mathcal{A}_2 + \mathcal{A}_3\right),\qquad \mathcal{A}_1 =
\frac{\Theta_+(h)}{2\pi} \lim_{r^\star\to0}
\!\!\!\int\limits_{|\vec{r} - \vec{r}'|>r^\star} \!\!\!
\mathrm{d}^2x' \int_0^h \mathrm{d}z
\frac{(2z^2-\rho^2)(h-z+a_0)}{\left( \rho^2 + z^2
\right)^{5/2}},\\
&\mathcal{A}_2 = -\frac{a_0^4}{8\pi}
\sum_{\substack{{\vec{\mu}}\\r_{\vec{n}\vec{m}}\neq0}}
\frac{1}{\rho_{{\vec{\nu}}{\vec{\mu}}}^3} \approx \frac{a_0^2}{8\pi}
\int_0^{2\pi} \frac{\mathrm{d}\alpha}{P} - \frac{a_0}{4},\quad
\mathcal{A}_3 = \frac{a_0^2}{8\pi} \int \mathrm{d}^2x'
\frac{2h^2-\rho^2}{\left( \rho^2 + h^2 \right)^{5/2}} \approx
\frac{a_0^2}{8\pi} \int_0^{2\pi}
\frac{P^2\mathrm{d}\alpha}{(P^2+h^2)^{3/2}} -
\frac{a_0^4}{4(a_0^2+h^2)^{3/2}}.\nonumber
\end{align}
Here $\rho = \sqrt{(x-x')^2 + (y-y')^2}$ and we used a local reference frame
\eqref{eq:loc-frame} and the Heaviside function $\Theta_+(x)$ takes the unit
values for any positive $x$ and zero values for $x\leq0$. The Heaviside
function is added here to fulfil the condition $\mathcal{A}_1\equiv 0$ in a 2D
case, when for $h=0$. There is a singularity in $\mathcal{A}_1$, due to the
nonintegrability of the kernel $K_z$ at $r_{\vec{n}\vec{m}}=0$. To regularize
it we use a method similar to the one in Ref.~\onlinecite{Akhiezer68}.
Specifically, we present $\mathcal{A}_1$ in the form $\mathcal{A}_1=
\widetilde{\mathcal{A}_1} - \mathcal{A}_0$. The coefficient
$\widetilde{\mathcal{A}_1}$ is a regular one:
\begin{equation*}
\widetilde{\mathcal{A}_1} = \frac{\Theta_+(h)}{2\pi} \int
\mathrm{d}^2x' \int_0^h \mathrm{d}z
\frac{(2z^2-\rho^2)(h-z+a_0)}{\left( \rho^2 + z^2 \right)^{5/2}}=
-h-a_0\Theta_+(h) + \frac{1}{2\pi} \int_0^{2\pi} \mathrm{d}\alpha
\left[ \sqrt{P^2+h^2} - P
 + \frac{a_0h}{\sqrt{P^2+h^2}}\right].
\end{equation*}
The singularity is inside the $\mathcal{A}_0$ term:
\begin{equation} \label{eq:A0}
\begin{split}
\mathcal{A}_0 &= \frac{\Theta_+(h)}{2\pi} \lim_{r^\star\to0} \!\!\!
\int\limits_{\substack{|\vec{r} - \vec{r}'|<r^\star\\{z=0,}\;z'>0}}
\!\!\!\!\!\!\! \mathrm{d}^2 x' \mathrm{d}z' \frac{(2{z'}^2 -
\rho^2)(h-z'+a_0)}{\left(\rho^2 + {z'}^2\right)^{5/2}} =
\frac{\Theta_+(h)}{2\pi}\Bigl[(h+a_0)I_1 - I_2 \Bigr],\\
I_1&= \lim_{r^\star\to0} \!\!\! \int\limits_{\substack{|\vec{r} -
\vec{r}'|<r^\star\\ z=0,\;z'>0}} \!\!\!\!\!\!\! \mathrm{d}^2 x'
\mathrm{d}z' \frac{2z'^2-\rho^2}{\left(\rho^2 + z'^2\right)^{5/2}} =
\lim_{r^\star\to0}
\!\!\!\!\! \int\limits_{\substack{|\vec{r} - \vec{r}'|<r^\star\\
z=0,\;z'>0}} \!\!\!\!\!\!\! \mathrm{d}^2 x' \mathrm{d}z'
\frac{\partial^2}{\partial {z'}^2}\frac{1}{|\vec{r}-\vec{r}'|}
=\frac13
\lim_{r^\star\to0} \!\!\!\!\! \int\limits_{\substack{|\vec{r} - \vec{r}'|<r^\star\\
z=0,\;z'>0}} \!\!\!\!\!\!\! \mathrm{d}^3x'\Delta\frac{1}{|\vec{r}-\vec{r}'|}\\
&=-\frac{4\pi}{3} \lim_{r^\star\to0}
\!\!\!\!\! \int\limits_{\substack{|\vec{r} - \vec{r}'|<r^\star\\
z=0,\;z'>0}} \!\!\!\!\!\!\! \mathrm{d}^3x'\delta(\vec{r} - \vec{r}')
=-\frac{2\pi}{3},\\
I_2&=\lim_{r^\star\to0} \!\!\! \int\limits_{\substack{|\vec{r} - \vec{r}'|<r^\star\\
z=0,\;z'>0}} \!\!\!\!\!\!\! \mathrm{d}^2 x' \mathrm{d}z'
\frac{z'(2z'^2-\rho^2)}{\left(\rho^2 + z'^2\right)^{5/2}}
=\frac{4\pi}{3}\lim_{r^\star\to0} \!\!\!
\int\limits_{\substack{|\vec{r} - \vec{r}'|<r^\star\\ z=0,\;z'>0}}
\!\!\!\!\!\!\! \mathrm{d}^3x'z'\delta(\vec{r} - \vec{r}')=0.
\end{split}
\end{equation}
Finally, $\mathcal{A}_0=-\left[h+a_0\Theta_+(h)\right]/3$ and the coefficient
of effective anisotropy $\mathcal{A}(x,y)$ takes a form \eqref{eq:A-fin}.

The coefficients $\bar{B}_{{\vec{\nu}}}$ and $\bar{C}_{{\vec{\nu}}}$ can be
calculated in the same way, starting from Eq.~\eqref{eq:B-details}:
\begin{align} \label{eq:B-and-C}
&\mathcal{B}(x,y) \equiv -\frac{a_0^4 e^{2\imath\chi}}{2\pi h}
\Bigl[ \bar{B}_{{\vec{\nu}}} - \imath \bar{C}_{{\vec{\nu}}}
\Bigr] = -\frac{3a_0^4}{4\pi h}
\sum_{\substack{{\vec{\mu}}\\r_{\vec{n}\vec{m}}\neq0}}
\rho_{{\vec{\nu}}{\vec{\mu}}}^2 e^{-2\imath
\alpha_{{\vec{\nu}}{\vec{\mu}}}} \sum_{n_z,m_z}
\frac{1}{\left(\rho_{{\vec{\nu}}{\vec{\mu}}}^2 +
z_{\vec{n}\vec{m}}^2\right)^{5/2}} = \frac{1}{h}\left( \mathcal{B}_1
+ \mathcal{B}_2 + \mathcal{B}_3\right),\\
&\mathcal{B}_1 = -\frac{3}{2\pi}\!\!\int\!\! \mathrm{d}^2x' \rho^2
e^{-2\imath\alpha}\!\!\! \int\limits_0^h \!\!\mathrm{d}z
\frac{h-z+a_0}{(\rho^2+z^2)^{5/2}} =\frac{1}{2\pi}\!\!\!
\int\limits_0^{2\pi}\!\!\mathrm{d}\alpha e^{-2\imath\alpha}\!\!
\Biggl[ P \!- \! \sqrt{P^2+h^2} + \frac{a_0 h}{\sqrt{P^2+h^2}}
- 2\left(h+a_0\right) \ln\frac{\sqrt{P^2+h^2}-h}{P} \Biggr],\nonumber \\
&\mathcal{B}_2 = -\frac{3 a_0^4}{8\pi}
\sum_{\substack{{\vec{\mu}}\\r_{\vec{n}\vec{m}}\neq0}}
\frac{e^{-2\imath\alpha}}{\rho_{{\vec{\nu}}{\vec{\mu}}}^3} \approx
\frac{3 a_0^2}{8\pi} \int_0^{2\pi}
\frac{e^{-2\imath\alpha}\mathrm{d}\alpha}{P},\qquad \mathcal{B}_3 =
-\frac{3a_0^2}{8\pi} \int \mathrm{d}^2x' \frac{\rho^2
e^{-2\imath\alpha}}{\left( \rho^2 + h^2 \right)^{5/2}} =
\frac{a_0^2}{8\pi} \int_0^{2\pi} \mathrm{d}\alpha
e^{-2\imath\alpha}\frac{3P^2 + 2h^2}{(P^2+h^2)^{3/2}}.\nonumber
\end{align}
Finally, the coefficient of effective anisotropy $\mathcal{B}(x,y)$ takes a
form \eqref{eq:B-fin}. As a result the dipolar energy
\eqref{eq:Energy-dipolar} can be expressed as \eqref{eq:E-dd-eff}.

Note that for the circular system one can obtain exact expressions for  the
coefficients $\mathcal{A}$ and $\mathcal{B}$. Let us first find the
coefficient $\mathcal{A}$. Assuming that $h\gg a_0$ (or equivalently
$a_0\to0$), one can rewrite the coefficient $\mathcal{A}$, see
Eq.~\eqref{eq:A-calculation}, as follows:
\begin{equation} \label{eq:A-circular}
\begin{split}
\mathcal{A}(\xi) &= \frac13 + \frac1{4\pi\varepsilon}
\Bigl[I_A(2\varepsilon) - I_A(0)\Bigr],\quad
I_A(x) = \int_0^{2\pi} \mathrm{d}\alpha \int_0^1 \frac{\xi'\mathrm{d}\xi'}{
\sqrt{\xi^2 + {\xi'}^2 + x^2 - 2\xi\xi'\cos\alpha}},\\
I_A(x) &=\frac{2}{\sqrt{x^2+(\xi +1)^2}}
\Bigl\{\left[x^2+(\xi +1)^2\right]
\text{E}(\mu)+\left[1-x^2-\xi^2\right] \text{K}(\mu) +
F_+(x) + F_-(x)\Bigr\}-2 \pi x,\\
F_\pm(x) &= x^2 \frac{\sqrt{x^2+\xi^2}\mp1}{\sqrt{x^2+\xi ^2}\pm \xi }
\Pi \left(\nu_\pm|\mu\right), \qquad \mu = \frac{4\xi}{x^2+(1+\xi)^2}, \quad
\nu_\pm = \frac{2\xi}{\xi\pm\sqrt{x^2+\xi^2}},
\end{split}
\end{equation}
where $\Pi(\nu_\pm|\mu)$ is the complete elliptic integral of the third kind.
\cite{Abramowitz64}.

To calculate the in-plane anisotropy coefficient $\mathcal{B}$, see
Eq.~\eqref{eq:B-and-C}, it is convenient to use the following relations
\begin{align} \label{eq:Stokes}
&\text{Re}\left[\mathcal{B} e^{-2\imath\chi}\right]
=-\frac{a_0^4}{2\pi h}\bar{B}_{{\vec{\nu}}}=-\frac{1}{2\pi h}
\int_0^h \mathrm{d}z (h-z)
I_z(x),\qquad I_z(x) = 3\int \mathrm{d}^2x'
\,\frac{(x-x')^2-(y-y')^2}{
\left( \rho^2 + z^2\right)^{5/2}}\\
&=\int \mathrm{d}^2x'\left(\frac{\partial^2}{\partial y\,\partial
y'}-\frac{\partial^2}{\partial x\,\partial x'}\right)\frac{1}{
\sqrt{\rho^2+ z^2}}
\equiv\int_\Omega \left[\vec{\nabla}'\times \vec{F}\right]\cdot
\mathrm{d}\vec{\sigma}
=\oint_{\partial\Omega} \vec{F}\cdot \mathrm{d}\vec{l'}, \quad \vec{F} = \vec{e}_z
\times \vec{\nabla}\frac{1}{\sqrt{(x-{x'})^2 + (y-{y'})^2+ z^2}}.\nonumber
\end{align}
For a circular system
$\mathrm{d}\vec{l'}=R\mathrm{d}\chi'\left(-\vec{e}_x\sin\chi' + \vec{e}_y
\cos\chi'\right)$, hence
\begin{equation*}
I_z(x) = R\!\! \int\limits_0^{2\pi}\!\!\mathrm{d}\chi'\!\!
\left[\frac{\partial}{\partial y}\sin\chi'\!-\!
\frac{\partial}{\partial x}\cos\chi'\right]
\!\! \frac1{\sqrt{r^2+R^2-2 rR \cos(\chi-\chi')+ z^2}}
=rR\cos(2\chi)\frac{\partial}{\partial r}\!
\left[\frac{1}{r} \! \int\limits_0^{2\pi}\!\!\frac{\cos\alpha\,\mathrm{d}\alpha}{
\sqrt{r^2+R^2-2 rR\cos\alpha+ z^2 }}\right].
\end{equation*}
Taking into account that $\text{Im}\mathcal{B}=0$ for the circular system, one
can calculate finally the effective in-plane anisotropy coefficient
$\mathcal{B}$ as follows:
\begin{equation} \label{eq:B-circular}
\begin{split}
&\mathcal{B}(\xi) = \frac{1}{2\pi\varepsilon}
\Bigl[I_B(2\varepsilon) - I_B(0)\Bigr],\quad I_B(x) =
c_1\text{K}(\mu)+c_2 \text{E}(\mu)+c_3
\Pi \left(\left.\frac{4\xi}{(1+\xi)^2}\right|\mu\right),\\
&c_1=\frac{2-2 x^2-\xi ^2-\left(x^2+\xi^2\right)^2}{3\xi^2\sqrt{x^2+(1+\xi)^2}},
\quad c_2=\frac{\left(x^2+\xi ^2-2\right)\sqrt{x^2+(1+\xi)^2}}{3 \xi ^2},\quad
c_3=\frac{x^2 (1-\xi )}{\xi ^2 (1+\xi)\sqrt{x^2+(\xi +1)^2}}.
\end{split}
\end{equation}
The dipolar energy $W_d$ [see Eq.~\eqref{eq:W-total-res}] for the disk--shaped
system can be presented in the form $W_d= W_d^0 + \widetilde{W_d}$, where
\begin{equation} \label{eq:Wd-disk}
\widetilde{W_d} = \frac{1}{R^2}\int\mathrm{d}^2x
\Bigl[\widetilde{\mathcal{A}}(r) + \mathcal{B}(r)\cos
2(\phi-\chi)\Bigr] \sin^2\theta
\end{equation}
and $W_d^0 = -2R^{-2}\int\mathrm{d}^2x \mathcal{A}(r)$ being the isotropic
part, the effective easy-plane anisotropy parameter $\widetilde{\mathcal{A}} =
3\mathcal{A}$.
\end{widetext}

\section{Configurational anisotropy for a half-plane and a square prism}
\label{sec:appendix-2geom} %

We start here with the problem for a half-plane. Consider the large scale
behavior of the dipolar energy, given by the in--plane effective anisotropy
$\mathcal{B}(x,y)$, see Eq.~\eqref{eq:A&B}. Straightforward calculations lead
to the effective anisotropy constant for the upper half-plane
\begin{equation} \label{eq:B-exact}
\begin{split}
\mathcal{B}(x,y) &\equiv \mathcal{B}(y_0) =
\frac{1}{2\pi}\int_{-\infty}^\infty \mathrm{d}x
y_{0}(y_{0}^2-x^2)\frac{\mathcal{F}(P,h)}{P^4}\\
&=\frac{y_{0}}{2\pi h}\ln\frac{y_{0}^2}{y_{0}^2 + h^2} + \frac{1}{\pi}
\arctan\frac{h}{y_{0}},
\end{split}
\end{equation}
where we choose the origin of the local reference frame at the boundary of the
domain, at $(x,y) = (0,0)$, $y_0$ denotes the distance from the boundary, and
$P=\sqrt{{x}^2 + y_0^2}$. One can see that $\mathcal{B}$ does not depend on
$x$, it takes only positive real values, hence $\arg\mathcal{B}=0$ for
\emph{any} distances $y_{0}$ from the boundary. This means that the in--plane
spin angle $\phi$ is always parallel to the half-plane edge. Using
Eqs.~\eqref{eq:B-vol}, \eqref{eq:F-vol} and \eqref{eq:B-exact} we found that
the main contribution to \eqref{eq:B-exact} is provided by the boundary domain
$x\in [-R_{0};R_{0}]$ with $R_{0}\sim\sqrt{y_{0}h}$. Since this domain
collapses to a point when $y_{0}\to0$, we conclude that for any geometry the
in-plane spin distribution is parallel to the boundary near the edge. If the
curvature radius of the sample boundary is larger than $R_{0}$, then spins are
parallel to the boundary over a distance smaller than $R_{0}^2/h$. One should
remember, that this conclusion is adequate for regions, where exchange
interaction has no principal influence.

\begin{figure}
\includegraphics[width=\columnwidth]{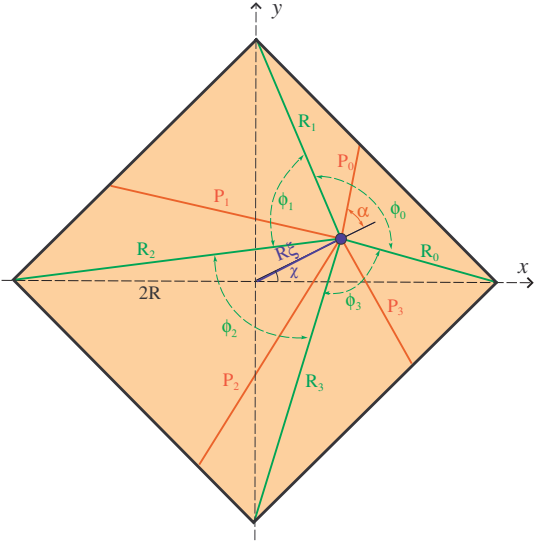}
\caption{(Color online) Arrangement of coordinates in the local reference
frame for the prism shaped particle.}
\label{fig:MobFrameSquare} %
\end{figure}

Let us consider now the configurational anisotropy for the square prism, which
has the diagonal $2R$, see Fig.~\ref{fig:MobFrameSquare}.
It is convenient to use the local reference frame in the same way as in
Sec.~\ref{sec:continuum}. The relative polar coordinates are defined as
follows:
\begin{equation}\label{eq:square_params}
\begin{split}
R_n=&R\sqrt{1+\xi^2-2\xi\cos\left(n\pi/2-\chi\right)},\\
\varphi_n=&\arccos\frac{R_n^2+R_{n+1}^2-2R^2}{2R_nR_{n+1}},\\
P_n=&\frac{R_{n}R_{n+1}}{R\sqrt{2}}\frac{\sin\varphi_n}{\cos\left(\alpha
+\chi-(2n+1)\pi/4\right)},
\end{split}
\end{equation}
where $\xi=\sqrt{x^2+y^2}/R$. Now we are able to compute magnetization
distribution on a large scale, which follows from the minimization condition
\eqref{eq:phi-min}. Straightforward calculations give
\begin{align}\label{eq:phi-min-1}
\phi &= \chi + \frac{\pi}{2} - \frac12 \text{Arg}\mathcal{B},\\
\label{eq:phi-min-2} %
\mathcal{B} &= \frac{1}{2\pi} \Biggl[
\int_{\psi_0-\varphi_0}^{\psi_0}e^{-2\imath\alpha}
\mathcal{F}(P_0,h)\mathrm{d}\alpha\nonumber\\
& + \sum\limits_{j=1}^3\int_{\psi_{j-1}}^{\psi_j}
e^{-2\imath\alpha}\mathcal{F}(P_j,h) \mathrm{d}\alpha \Biggr],\\
\label{eq:phi-min-3} %
\psi_j &= \psi_0+\sum_{i=1}^j\varphi_i,\\
\label{eq:phi-min-4} %
\psi_0 &= \frac{3\pi}{4} - \chi - \arcsin\left(\frac{R_0
\sin\varphi_0}{R\sqrt{2}}\right),
\end{align}
where $\mathcal F(P_i,h)$ is defined by \eqref{eq:F-vol}.


\begin{thebibliography}{32}
\expandafter\ifx\csname natexlab\endcsname\relax\def\natexlab#1{#1}\fi
\expandafter\ifx\csname bibnamefont\endcsname\relax
  \def\bibnamefont#1{#1}\fi
\expandafter\ifx\csname bibfnamefont\endcsname\relax
  \def\bibfnamefont#1{#1}\fi
\expandafter\ifx\csname citenamefont\endcsname\relax
  \def\citenamefont#1{#1}\fi
\expandafter\ifx\csname url\endcsname\relax
  \def\url#1{\texttt{#1}}\fi
\expandafter\ifx\csname urlprefix\endcsname\relax\def\urlprefix{URL }\fi
\providecommand{\bibinfo}[2]{#2}
\providecommand{\eprint}[2][]{\url{#2}}

\bibitem[{\citenamefont{Hubert and Sch{\" a}fer}(1998)}]{Hubert98}
\bibinfo{author}{\bibfnamefont{A.}~\bibnamefont{Hubert}} \bibnamefont{and}
  \bibinfo{author}{\bibfnamefont{R.}~\bibnamefont{Sch{\" a}fer}},
  \emph{\bibinfo{title}{Magnetic domains}}
  (\bibinfo{publisher}{Springer--Verlag}, \bibinfo{address}{Berlin},
  \bibinfo{year}{1998}).

\bibitem[{\citenamefont{Skomski}(2003)}]{Skomski03}
\bibinfo{author}{\bibfnamefont{R.}~\bibnamefont{Skomski}},
  \bibinfo{journal}{J.~Phys.} \textbf{\bibinfo{volume}{C 15}},
  \bibinfo{pages}{R841} (\bibinfo{year}{2003}),
  \urlprefix\url{http://www.iop.org/EJ/abstract/0953-8984/15/20/202/}.

\bibitem[{\citenamefont{Bader}(2006)}]{Bader06}
\bibinfo{author}{\bibfnamefont{S.~D.} \bibnamefont{Bader}},
  \bibinfo{journal}{Reviews of Modern Physics} \textbf{\bibinfo{volume}{78}},
  \bibinfo{eid}{1} (pages~\bibinfo{numpages}{15}) (\bibinfo{year}{2006}),
  \urlprefix\url{http://link.aps.org/abstract/RMP/v78/p1}.

\bibitem[{\citenamefont{Cowburn}(2002)}]{Cowburn02}
\bibinfo{author}{\bibfnamefont{R.~P.} \bibnamefont{Cowburn}},
  \bibinfo{journal}{J.~Magn. Magn. Mater.} \textbf{\bibinfo{volume}{242-245}},
  \bibinfo{pages}{505} (\bibinfo{year}{2002}),
  \urlprefix\url{http://www.sciencedirect.com/science/article/B6TJJ-44NM486-19%
/2/d76afadecc91907bb34369038cc8368e}.

\bibitem[{\citenamefont{Brown}(1963)}]{Brown63}
\bibinfo{author}{\bibfnamefont{W.~F.} \bibnamefont{Brown}, \bibfnamefont{Jr.}},
  \emph{\bibinfo{title}{Micromagnetism}} (\bibinfo{publisher}{Wiley, New York},
  \bibinfo{year}{1963}).

\bibitem[{\citenamefont{Aharoni}(1996)}]{Aharoni96}
\bibinfo{author}{\bibfnamefont{A.}~\bibnamefont{Aharoni}},
  \emph{\bibinfo{title}{Introduction to the theory of ferromagnetism}}
  (\bibinfo{publisher}{Oxford University Press}, \bibinfo{year}{1996}).

\bibitem[{\citenamefont{Stoner and Wohlfarth}(1948)}]{Stoner48}
\bibinfo{author}{\bibfnamefont{E.}~\bibnamefont{Stoner}} \bibnamefont{and}
  \bibinfo{author}{\bibfnamefont{E.}~\bibnamefont{Wohlfarth}},
  \bibinfo{journal}{Philosophical Transactions of the Royal Society of London.
  Series A, Mathematical and Physical Sciences (1934-1990)}
  \textbf{\bibinfo{volume}{240}}, \bibinfo{pages}{599} (\bibinfo{year}{1948}),
  \urlprefix\url{http://www.journals.royalsoc.ac.uk/openurl.asp?genre=article&%
id=V616652U27003458}.

\bibitem[{\citenamefont{Usov and Peschany}(1992)}]{Usov92}
\bibinfo{author}{\bibfnamefont{N.~A.} \bibnamefont{Usov}} \bibnamefont{and}
  \bibinfo{author}{\bibfnamefont{S.~E.} \bibnamefont{Peschany}},
  \bibinfo{journal}{J.~Magn. Magn. Mater.} \textbf{\bibinfo{volume}{110}},
  \bibinfo{pages}{L1} (\bibinfo{year}{1992}).

\bibitem[{\citenamefont{Cowburn et~al.}(1998)\citenamefont{Cowburn, Adeyeye,
  and Welland}}]{Cowburn98}
\bibinfo{author}{\bibfnamefont{R.~P.} \bibnamefont{Cowburn}},
  \bibinfo{author}{\bibfnamefont{A.~O.} \bibnamefont{Adeyeye}},
  \bibnamefont{and} \bibinfo{author}{\bibfnamefont{M.~E.}
  \bibnamefont{Welland}}, \bibinfo{journal}{Phys. Rev. Lett.}
  \textbf{\bibinfo{volume}{81}}, \bibinfo{pages}{5414} (\bibinfo{year}{1998}),
  \urlprefix\url{http://link.aps.org/abstract/PRL/v81/p5414}.

\bibitem[{\citenamefont{Ivanov and Tartakovskaya}(2004)}]{Ivanov04}
\bibinfo{author}{\bibfnamefont{B.~A.} \bibnamefont{Ivanov}} \bibnamefont{and}
  \bibinfo{author}{\bibfnamefont{E.~V.} \bibnamefont{Tartakovskaya}},
  \bibinfo{journal}{JETP} \textbf{\bibinfo{volume}{98}}, \bibinfo{pages}{1015}
  (\bibinfo{year}{2004}).

\bibitem[{\citenamefont{Usov and Peschany}(1994)}]{Usov94}
\bibinfo{author}{\bibfnamefont{N.~A.} \bibnamefont{Usov}} \bibnamefont{and}
  \bibinfo{author}{\bibfnamefont{S.~E.} \bibnamefont{Peschany}},
  \bibinfo{journal}{Fiz. Met. Metal.} \textbf{\bibinfo{volume}{78}},
  \bibinfo{pages}{13} (\bibinfo{year}{1994}), \bibinfo{note}{(in russian)}.

\bibitem[{\citenamefont{Kireev and Ivanov}(2003)}]{Kireev03}
\bibinfo{author}{\bibfnamefont{V.~E.} \bibnamefont{Kireev}} \bibnamefont{and}
  \bibinfo{author}{\bibfnamefont{B.~A.} \bibnamefont{Ivanov}},
  \bibinfo{journal}{Phys. Rev. B} \textbf{\bibinfo{volume}{68}},
  \bibinfo{eid}{104428} (pages~\bibinfo{numpages}{9}) (\bibinfo{year}{2003}),
  \urlprefix\url{http://link.aps.org/abstract/PRB/v68/e104428}.

\bibitem[{\citenamefont{Tchernyshyov and Chern}(2005)}]{Tchernyshyov05}
\bibinfo{author}{\bibfnamefont{O.}~\bibnamefont{Tchernyshyov}}
  \bibnamefont{and} \bibinfo{author}{\bibfnamefont{G.-W.} \bibnamefont{Chern}},
  \bibinfo{journal}{Phys. Rev. Lett.} \textbf{\bibinfo{volume}{95}},
  \bibinfo{eid}{197204} (pages~\bibinfo{numpages}{4}) (\bibinfo{year}{2005}),
  \urlprefix\url{http://link.aps.org/abstract/PRL/v95/e197204}.

\bibitem[{\citenamefont{Cowburn and Welland}(1998)}]{Cowburn98a}
\bibinfo{author}{\bibfnamefont{R.~P.} \bibnamefont{Cowburn}} \bibnamefont{and}
  \bibinfo{author}{\bibfnamefont{M.~E.} \bibnamefont{Welland}},
  \bibinfo{journal}{Phys. Rev. B} \textbf{\bibinfo{volume}{58}},
  \bibinfo{pages}{9217} (\bibinfo{year}{1998}),
  \urlprefix\url{http://link.aps.org/abstract/PRB/v58/p9217}.

\bibitem[{\citenamefont{Gioia and James}(1997)}]{Gioia97}
\bibinfo{author}{\bibfnamefont{G.}~\bibnamefont{Gioia}} \bibnamefont{and}
  \bibinfo{author}{\bibfnamefont{R.~D.} \bibnamefont{James}},
  \bibinfo{journal}{Proc. R. Soc. Lond. A} \textbf{\bibinfo{volume}{453}},
  \bibinfo{pages}{213} (\bibinfo{year}{1997}),
  \urlprefix\url{http://www.journals.royalsoc.ac.uk/openurl.asp?genre=article&%
id=doi:10.1098/rspa.1997.0013}.

\bibitem[{\citenamefont{Desimone}(1995)}]{Desimone95}
\bibinfo{author}{\bibfnamefont{A.}~\bibnamefont{Desimone}},
  \bibinfo{journal}{Meccanica} \textbf{\bibinfo{volume}{30}},
  \bibinfo{pages}{591} (\bibinfo{year}{1995}),
  \urlprefix\url{http://dx.doi.org/10.1007/BF01557087}.

\bibitem[{\citenamefont{Desimone et~al.}(2002)\citenamefont{Desimone, Kohn,
  M{\"u}ller, and Otto}}]{Desimone02}
\bibinfo{author}{\bibfnamefont{A.}~\bibnamefont{Desimone}},
  \bibinfo{author}{\bibfnamefont{R.~V.} \bibnamefont{Kohn}},
  \bibinfo{author}{\bibfnamefont{S.}~\bibnamefont{M{\"u}ller}},
  \bibnamefont{and} \bibinfo{author}{\bibfnamefont{F.}~\bibnamefont{Otto}},
  \bibinfo{journal}{Communications on Pure and Applied Mathematics}
  \textbf{\bibinfo{volume}{55}}, \bibinfo{pages}{1408} (\bibinfo{year}{2002}),
  \urlprefix\url{http://dx.doi.org/10.1002/cpa.3028}.

\bibitem[{\citenamefont{Kohn and Slastikov}(2005{\natexlab{a}})}]{Kohn05}
\bibinfo{author}{\bibfnamefont{R.}~\bibnamefont{Kohn}} \bibnamefont{and}
  \bibinfo{author}{\bibfnamefont{V.}~\bibnamefont{Slastikov}},
  \bibinfo{journal}{Proc. R. Soc. A} \textbf{\bibinfo{volume}{461}},
  \bibinfo{pages}{143} (\bibinfo{year}{2005}{\natexlab{a}}),
  \urlprefix\url{http://dx.doi.org/10.1098/rspa.2004.1342}.

\bibitem[{\citenamefont{Kohn and Slastikov}(2005{\natexlab{b}})}]{Kohn05a}
\bibinfo{author}{\bibfnamefont{R.~V.} \bibnamefont{Kohn}} \bibnamefont{and}
  \bibinfo{author}{\bibfnamefont{V.~V.} \bibnamefont{Slastikov}},
  \bibinfo{journal}{Archive for Rational Mechanics and Analysis}
  \textbf{\bibinfo{volume}{178}}, \bibinfo{pages}{227}
  (\bibinfo{year}{2005}{\natexlab{b}}),
  \urlprefix\url{http://dx.doi.org/10.1007/s00205-005-0372-7}.

\bibitem[{\citenamefont{Moser}(2004)}]{Moser04}
\bibinfo{author}{\bibfnamefont{R.}~\bibnamefont{Moser}},
  \bibinfo{journal}{Archive for Rational Mechanics and Analysis}
  \textbf{\bibinfo{volume}{174}}, \bibinfo{pages}{267} (\bibinfo{year}{2004}),
  \urlprefix\url{http://dx.doi.org/10.1007/s00205-004-0329-2}.

\bibitem[{\citenamefont{Kurzke}(2006)}]{Kurzke06}
\bibinfo{author}{\bibfnamefont{M.}~\bibnamefont{Kurzke}},
  \bibinfo{journal}{Calculus of Variations and Partial Differential Equations}
  \textbf{\bibinfo{volume}{26}}, \bibinfo{pages}{1} (\bibinfo{year}{2006}),
  \urlprefix\url{http://dx.doi.org/10.1007/s00526-005-0331-z}.

\bibitem[{\citenamefont{Ivanov and Zaspel}(2005)}]{Ivanov05}
\bibinfo{author}{\bibfnamefont{B.~A.} \bibnamefont{Ivanov}} \bibnamefont{and}
  \bibinfo{author}{\bibfnamefont{C.~E.} \bibnamefont{Zaspel}},
  \bibinfo{journal}{Phys. Rev. Lett.} \textbf{\bibinfo{volume}{94}},
  \bibinfo{pages}{027205} (\bibinfo{year}{2005}),
  \urlprefix\url{http://link.aps.org/abstract/PRL/v94/e027205}.

\bibitem[{\citenamefont{Akhiezer et~al.}(1968)\citenamefont{Akhiezer,
  Bar'yakhtar, and Peletminski\u{\i}}}]{Akhiezer68}
\bibinfo{author}{\bibfnamefont{A.~I.} \bibnamefont{Akhiezer}},
  \bibinfo{author}{\bibfnamefont{V.~G.} \bibnamefont{Bar'yakhtar}},
  \bibnamefont{and} \bibinfo{author}{\bibfnamefont{S.~V.}
  \bibnamefont{Peletminski\u{\i}}}, \emph{\bibinfo{title}{Spin waves}}
  (\bibinfo{publisher}{North--Holland}, \bibinfo{address}{Amsterdam},
  \bibinfo{year}{1968}).

\bibitem[{\citenamefont{L{\'e}vy}(2001)}]{Levy01}
\bibinfo{author}{\bibfnamefont{J.-C.~S.} \bibnamefont{L{\'e}vy}},
  \bibinfo{journal}{Phys. Rev. B} \textbf{\bibinfo{volume}{63}},
  \bibinfo{pages}{104409} (\bibinfo{year}{2001}),
  \urlprefix\url{http://link.aps.org/abstract/PRB/v63/e104409}.

\bibitem[{\citenamefont{Abramowitz and Stegun}(1964)}]{Abramowitz64}
\bibinfo{author}{\bibfnamefont{M.}~\bibnamefont{Abramowitz}} \bibnamefont{and}
  \bibinfo{author}{\bibfnamefont{I.~A.} \bibnamefont{Stegun}},
  \emph{\bibinfo{title}{Handbook of mathematical functions with formulas,
  graphs, and mathematical tables}} (\bibinfo{publisher}{Dover},
  \bibinfo{address}{New York}, \bibinfo{year}{1964}), \bibinfo{edition}{ninth
  dover printing, tenth gpo printing} ed., ISBN \bibinfo{isbn}{0-486-61272-4}.

\bibitem[{\citenamefont{Joseph}(1966)}]{Joseph66}
\bibinfo{author}{\bibfnamefont{R.~I.} \bibnamefont{Joseph}},
  \bibinfo{journal}{J.~Appl. Phys.} \textbf{\bibinfo{volume}{37}},
  \bibinfo{pages}{4639} (\bibinfo{year}{1966}),
  \urlprefix\url{http://link.aip.org/link/?JAP/37/4639/1}.

\bibitem[{\citenamefont{Aharoni}(1990)}]{Aharoni90}
\bibinfo{author}{\bibfnamefont{A.}~\bibnamefont{Aharoni}},
  \bibinfo{journal}{J.~Appl. Phys.} \textbf{\bibinfo{volume}{68}},
  \bibinfo{pages}{2892} (\bibinfo{year}{1990}),
  \urlprefix\url{http://link.aip.org/link/?JAP/68/2892/1}.

\bibitem[{\citenamefont{Wysin}(1994)}]{Wysin94}
\bibinfo{author}{\bibfnamefont{G.~M.} \bibnamefont{Wysin}},
  \bibinfo{journal}{Phys. Rev. B} \textbf{\bibinfo{volume}{49}},
  \bibinfo{pages}{8780} (\bibinfo{year}{1994}),
  \urlprefix\url{http://link.aps.org/abstract/PRB/v49/p8780}.

\bibitem[{\citenamefont{Ivanov and Sheka}(1995)}]{Ivanov95b}
\bibinfo{author}{\bibfnamefont{B.~A.} \bibnamefont{Ivanov}} \bibnamefont{and}
  \bibinfo{author}{\bibfnamefont{D.~D.} \bibnamefont{Sheka}},
  \bibinfo{journal}{Low Temp. Phys.} \textbf{\bibinfo{volume}{21}},
  \bibinfo{pages}{881} (\bibinfo{year}{1995}),
  \urlprefix\url{http://link.aip.org/link/?LTP/21/881/1}.

\bibitem[{\citenamefont{Kravchuk et~al.}(2007)\citenamefont{Kravchuk, Sheka,
  and Gaididei}}]{Kravchuk07}
\bibinfo{author}{\bibfnamefont{V.~P.} \bibnamefont{Kravchuk}},
  \bibinfo{author}{\bibfnamefont{D.~D.} \bibnamefont{Sheka}}, \bibnamefont{and}
  \bibinfo{author}{\bibfnamefont{Y.~B.} \bibnamefont{Gaididei}},
  \bibinfo{journal}{J.~Magn. Magn. Mater.} \textbf{\bibinfo{volume}{310}},
  \bibinfo{pages}{116} (\bibinfo{year}{2007}),
  \urlprefix\url{http://www.sciencedirect.com/science/article/B6TJJ-4KSSV4X-1/%
2/c32bf9c25779482f7e04cdbecfeb9a3d}.

\bibitem[{\citenamefont{H{\"o}llinger et~al.}(2003)\citenamefont{H{\"o}llinger,
  Killinger, and Krey}}]{Hoellinger03}
\bibinfo{author}{\bibfnamefont{R.}~\bibnamefont{H{\"o}llinger}},
  \bibinfo{author}{\bibfnamefont{A.}~\bibnamefont{Killinger}},
  \bibnamefont{and} \bibinfo{author}{\bibfnamefont{U.}~\bibnamefont{Krey}},
  \bibinfo{journal}{J.~Magn. Magn. Mater.} \textbf{\bibinfo{volume}{261}},
  \bibinfo{pages}{178} (\bibinfo{year}{2003}),
  \urlprefix\url{http://www.sciencedirect.com/science/article/B6TJJ-47F1D5K-1/%
2/02a9f675dac9cbb50d643664d50b98be}.

\bibitem[{OOM()}]{OOMMF}
\emph{\bibinfo{title}{The {O}bject {O}riented {M}icro{M}agnetic {F}ramework}},
  \bibinfo{note}{developed by M. J. Donahue and D. Porter mainly, from NIST. We
  used the 3D version of the 1.2$\alpha$2 release},
  \urlprefix\url{http://math.nist.gov/oommf/}.

\end{thebibliography}

\end{document}